\documentclass[aps,prd,onecolumn,showpacs,groupedaddress]{revtex4-2}
\usepackage{graphicx}% Include figure files
\usepackage{dcolumn}% Align table columns on decimal point
\usepackage{bm}% bold math
\usepackage{color}
\usepackage{longtable}
\usepackage{supertabular}
\usepackage[T1]{fontenc}
\usepackage{epstopdf}
\usepackage{amsmath}
\usepackage{amssymb}
\usepackage{setspace}
\usepackage{multirow}
\usepackage{booktabs}
\usepackage[section]{placeins}
\usepackage{mathrsfs}
\usepackage{hyperref}

\makeatletter

\newcommand{\Rmnum}[1]{\expandafter\@slowromancap\romannumeral #1@} 
\newtheorem{prop}{Proposition}
\newtheorem{coro}{Corollary}

\makeatother

\begin{document}
\title{On second-order combinatorial algebraic time-delay interferometry}

%\author{the authors}
\author{Wei-Liang Qian\textsuperscript{2,1,3}}\email[E-mail: ]{wlqian@usp.br}
\author{Pan-Pan Wang\textsuperscript{4}}\email[E-mail: ]{ppwang@hust.edu.cn}
\author{Zhang-Qi Wu\textsuperscript{4}}
\author{Cheng-Gang Shao\textsuperscript{4}}
\author{Bin Wang\textsuperscript{1,5}}
\author{Rui-Hong Yue\textsuperscript{1}}

\affiliation{$^{1}$ Center for Gravitation and Cosmology, College of Physical Science and Technology, Yangzhou University, Yangzhou 225009, China}
\affiliation{$^{2}$ Escola de Engenharia de Lorena, Universidade de S\~ao Paulo, 12602-810, Lorena, SP, Brazil}
\affiliation{$^{3}$ Faculdade de Engenharia de Guaratinguet\'a, Universidade Estadual Paulista, 12516-410, Guaratinguet\'a, SP, Brazil}
\affiliation{$^{4}$ MOE Key Laboratory of Fundamental Physical Quantities Measurement, Hubei Key Laboratory of Gravitation and Quantum Physics, PGMF, and School of Physics, Huazhong University of Science and Technology, Wuhan 430074, China}
\affiliation{$^{5}$ School of Aeronautics and Astronautics, Shanghai Jiao Tong University, Shanghai 200240, China}

\begin{abstract}
Inspired by the combinatorial algebraic approach proposed by Dhurandhar {\it et al.}, we propose two novel classes of second-generation time-delay interferometry (TDI) solutions and their further generalization.
The primary strategy of the algorithm is to enumerate specific types of residual laser frequency noise associated with second-order commutators in products of time-displacement operators.
The derivations are based on analyzing the delay time residual when expanded in time derivatives of the armlengths order by order.
It is observed that the solutions obtained by such a scheme are primarily captured by the geometric TDI approach and therefore possess an intuitive interpretation.
Nonetheless, the fully-symmetric Sagnac and Sagnac-inspired combinations inherit the properties from the original algebraic approach, and subsequently lie outside of the scope of geometric TDI.
We explicitly show that novel solutions, distinct from existing ones in terms of both algebraic structure and sensitivity curve, are encountered.
Moreover, at its lowest order, the solution is furnished by commutators of relatively compact form. 
Besides the original Michelson-type solution, we elaborate on other types of solutions such as the Monitor, Beacon, Relay, Sagnac, fully-symmetric Sagnac, and Sagnac-inspired ones.
The average response functions, residual noise power spectral density, and sensitivity curves are evaluated for the obtained solutions.
Also, the relations between the present scheme and other existing algorithms are discussed.

\end{abstract}
%\pacs{ 03.75.Dg, 06.30.Gv, 37.25. + k, 91.10.Pp}

%\date{\today}
\date{Oct. 1st, 2022}

\maketitle

\section{Introduction}\label{sec1}

The TDI algorithm was first introduced by Tinto \emph{et al.}~\cite{TDI-1} to suppress laser frequency noise in space-borne gravitational wave detectors~\cite{LISA-1, TianQin-1, Taiji-1}.
The solution is formulated using a proper combination of the delayed science data stream, effectively constructing a virtual equal-arm interferometer~\cite{Overview-1}.
Analogic to the original Michelson interferometer, most TDI solutions possess an intuitive geometric interpretation, giving rise to the so-called geometric TDI approach~\cite{Geo-TDI-1}.
From the algebraic perspective, the first-generation TDI~\cite{TDI1-1, TDI1.5-1}, associated with rigid armlengths, the solution space is a polynomial ring $\mathscr{R}$ in three~\cite{TA-1} and six~\cite{TA-2} variables over the rational numbers.
In particular, the problem can be reformulated to solve for the first {\it module of syzygies} of a left ideal of the ring~\cite{TA-1, TA-2} using the notion of Groebner basis~\cite{groebner} in computational algebra.

On the other hand, the second-generation TDI takes into account the nonrigid rotation of the three-spacecraft constellation~\cite{TDI2-1}.
While expressing the residual noise as an expansion in terms of time derivatives of the armlengths, the cancelation scheme is effectively truncated at the second order.
As a result, it provides a higher precision when compared to its first-generation counterpart.
From the algebraic perspective, the time-delay operators, which constitute the variables of the polynomial, can no longer be considered commutative.
The derivations of the second-generation TDI solutions are not straightforward.
In practice, the geometric TDI, a method of exhaustion, is employed chiefly to seek feasible TDI combinations by enumerating all possible close trajectories in the space-time diagram~\cite{Geo-TDI-1, Olaf, Geo-sister}.
Nonetheless, the solution space of the geometric TDI rapidly grows by $3^n$, where $n$ is the number of links, and therefore, the approach is computationally expensive at higher orders.
Also, its relevant solution space is somewhat restrictive since a feasible solution demands that successive transmissions of laser signals must be physically associated with neighboring links.
For instance, it is well-known that the fully symmetric Sagnac TDI solutions lie beyond such a solution space.

In the framework of the second-generation TDI, the use of the algebraic approach is somewhat restrictive owing to its non-commutative nature.
A notable exception is the combinatorial algebraic algorithm proposed by Dhurandhar \emph{et al.}~\cite{D2010}.
The original scheme mainly aimed for the Michelson-type scenario where one arm of the detector becomes temporarily dysfunctional.
Recently, the scheme has been extended to deal with a broadened selection of second-generation TDI combinations~\cite{cTDI2gen}.
Mathematically, the TDI solution is furnished by the kernel of the following homomorphism associated with a polynomial ring $\mathscr{R}$ in four variables 
\begin{equation}
\varphi: \mathscr{R}^2\to \mathscr{R} .
\end{equation}
The proposed algorithm is based on the properties of a specific commutator between two monomials defined by the products of particular time-displacement operators.
The algorithm essentially resides in the following three propositions.
First, there is a specific class of commutators that vanishes when one only considers the first-order contributions regarding the time derivatives of the armlengths.
Second, these commutators belong to the solution space of the second-generation TDI combinations.
Last but not least, they are identical to the residual laser frequency noise of the corresponding TDI solution.
In other words, by adequately enumerating (e.g., in the lexicographic order) such commutators, one manages to construct specific classes of TDI solutions systematically.
The crucial feature of the extended algorithm proposed by some of us~\cite{cTDI2gen} is to include time-advance operators into the existing propositions.
Subsequently, the approach can be applied to seek TDI solutions of the Monitor, Beacon, Relay, Sagnac, and fully symmetric Sagnac types.
Moreover, a novel set of Sagnac-inspired solutions was derived, which cannot be straightforwardly obtained using the geometric TDI. 

Compared to the method of exhaustion, an algebraic approach is beneficial because of its computational efficiency.
Moreover, it scrutinizes the solution space and potentially furnishes us with a better understanding of the relevant module of the non-commutative ring.
Therefore, further exploration regarding the relationship between the commutators of the time-displacement operators and second-generation TDI solutions is a worthy topic.
The present study is motivated by the above considerations.
It further generalizes the combinatorial algebraic approach to include two novel classes of second-generation TDI solutions.
This is achieved by elaborating on two forms of residual laser noise associated with second-order commutators between polynomials in time-translation operators.
As elaborated below, the associated residual laser noise manifestly vanishes regarding the contributions of first-order in the time derivatives of the armlengths.
The first class of TDI solutions involves the commutator between a polynomial and an arbitrary commutator.
In particular, the vanishing condition for the commutator imposed for the original algorithm is lifted: two monomials of equal length are replaced by two arbitrary polynomials. 
The second class is featured by the products of two polynomials in time-translation operators as its variables.
We show that the derived second-generation TDI solutions are not included in the original algebraic approach or its recent generalization.
While compared with those obtained before, the lowest-order solutions are furnished by commutators of a relatively compact form. 
Also, some solutions do not possess a straightforward geometric TDI interpretation.

The remainder of this letter is organized as follows. 
In Sec.~\ref{sec2}, we briefly review the TDI algorithm and present the utilized notations and conventions.
The combinatorial approach proposed by Dhurandhar {\it et al.} and its recent generalization are also revisited.
In Sec.~\ref{sec3}, we present the two novel classes of second-order TDI solutions.
We elaborate on the forms of the residual laser noise and the procedure to derive the corresponding TDI coefficients.
Subsequently, in Sec.~\ref{sec4}, we explore various classes of solutions, namely, the Michelson, Monitor, Beacon, Relay, Sagnac, fully-symmetric Sagnac, and Sagnac-inspired ones.
The average response functions, residual noise power spectral density, and the sensitivity curves of the obtained novel solutions are evaluated.
The concluding remarks are given in the last section. 
The complementary mathematical derivations will be relegated to the Appendices~\ref{appA},~\ref{appB}, and~\ref{appC}.

\section{TDI algorithm and the first-order combinatorial algebraic approach}\label{sec2}

\begin{figure}[h]
\includegraphics[width=0.50\textwidth]{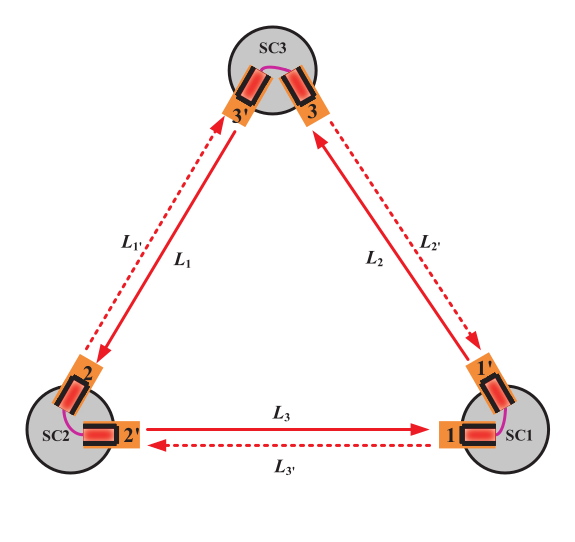}
\caption{\label{fig1} 
A schematic diagram of the three-spacecraft constellation for the space-borne gravitational wave detector.}
\end{figure}
\FloatBarrier

As shown in Fig.~\ref{fig1}, a typical space-based gravitational wave detector comprises a three-spacecraft constellation~\cite{Overview-1}.
The armlength sitting on the opposite side of the spacecraft $i$ (SC$i$) is denoted by $L_i$ (and $L_{i'}$) with $i=1, 2, 3$ in the counterclockwise (and clockwise) direction.
On each spacecraft, two lasers are installed on the corresponding optical benches, labeled by $i$ and $i'$.
Three types of data streams, namely, the science data stream $s_{i (i')}$, test mass data stream $\epsilon _{i (i')}$, and reference data stream $\tau _{i (i')}$, are recorded by the phasemeters.
The science data streams are the ones that carry the essential information on the gravitational waves, which triggers additional beat notes in the interference pattern.
By the standard procedure of the TDI algorithm, the data streams are post-processed offline.
The test mass and reference data streams are utilized to eliminate the optical bench motion noise.
Following the standard procedure of the TDI algorithm, the optical bench motion noise can be eliminated, and two local lasers are eﬀectively connected by intra-spacecraft phase locking~\cite{TC-1}.
The resultant observables read
	\begin{align}\label{eta}
		{\eta_i}(t) &= {H_i}(t) + {{D}_{i - 1}}{p_{i + 1}}(t) - {p_i}(t) + \nu_{(i+1)^{'}} {{\vec n}_{i - 1}}\left[ {{{D}_{i - 1}}{{\dot{\vec \delta }}_{(i + 1)'}}(t) - {{\dot {\vec \delta }}_i}(t)} \right] + N_i^{{opt}}(t) ,\notag\\
		{\eta_{i'}}(t) &= {H_{i'}}(t) + {{D}_{(i + 1)'}}{p_{i - 1}}(t) - {p_i}(t) + \nu_{i-1}{{\vec n}_{i + 1}} \cdot \left[ {{{\dot{\vec \delta }}_{i'}}(t) - {{D}_{(i + 1)'}}{{\dot{\vec{\delta} }}_{i - 1}}(t)} \right] + N_{i'}^{{opt}}(t) ,
	\end{align}
where the laser frequency noise is denoted by $p_i$, $\dot{\vec{\delta}}_{i, (i')}$ represents the test mass noise, $N_{i (i')}^{opt}$ gives the shot noise,
$H_{i, (i')}$ represent the gravitational wave signals, $D_{i (i')}$ are the time-delay operators along the related armlengths, and $\nu_{i (i')}$ are the laser's frequency.
In terms of the observables $\eta_{i, (i')}$, a valid TDI solution~\cite{Overview-1} aim to eliminate the laser frequency noise $p_i$ by the combination
\begin{equation}\label{TDI}
	\mathrm{TDI}=\sum_{i=1,2,3} (q_{i} \eta_{i} + q_{i^{'}} \eta_{i'}) ,
\end{equation}
where $q_{i}$ and $q_{i'}$ are polynomials in the six time-delay operators.
By focusing on the laser frequency noise and explicitly demanding the coefficients before individual $p_{i}$ vanish, the above equation gives
\begin{equation}\label{TDIeq}
	\begin{aligned}
		&q_{1}+q_{1^{\prime}}-q_{2^{\prime}} {D}_{3^{\prime}}-q_{3} {D}_{2}=0,\\
		&q_{2}+q_{2^{\prime}}-q_{3^{\prime}} {D}_{1^{\prime}}-q_{1} {D}_{3}=0, \\
		&q_{3}+q_{3^{\prime}}-q_{1^{\prime}} {D}_{2^{\prime}}-q_{2} {D}_{1}=0.
	\end{aligned}
\end{equation}

In particular, the Michelson-type TDI solution corresponds to when one does not use the data from the link opposite the recombining spacecraft.
Without loss of generality, one assumes that the link connecting SC2 and SC3 is not utilized.
Therefore one has $\eta_{2}=\eta_{{3}^{'}}=0$, or equivalently,
\begin{equation}\label{MC}
	q_{2}=q_{3^{'}}=0 .
\end{equation}
Substituting Eq.~\eqref{MC} into Eq.~\eqref{TDIeq}, and using two of the equations to eliminate $q_{2'}$ and $q_3$, one finds
\begin{equation}\label{MTDI}
	q_{1}(1-D_{33^{'}})+q_{1^{'}}(1-D_{2^{'}2})=0.
\end{equation}
The above equation can be rewritten in a generalized form
\begin{equation}\label{nMTDIgen}
	\alpha(1-a)+\beta(1-b)=0 ,
\end{equation}
where $a \equiv D_{33^{'}}$ and $b \equiv D_{2^{'}2}$, and the unknown coefficients $\alpha$ and $\beta$ are polynomials in $a$ and $b$.

The combinatorial approach consists of the following key ingredients~\cite{D2010, cTDI2gen}.
On the one hand, we focus on a particular class of commutators.
Such commutators vanish from the perspective of second-generation TDI; namely, the residuals are the second and higher-order terms regarding the time derivatives of the armlengths.
In other words, if these commutators can be mapped to the residual laser noise of some specific second-generation TDI combinations, they can be employed to furnish TDI solutions.
As illustrated below in Eq.~\eqref{M16V2}, particular examples indicate that the scheme is plausible.
In Refs.~\cite{D2010, cTDI2gen}, the above statement was generally shown to hold.
On the other hand, one can elaborate an algorithm to enumerate such commutators systematically and, subsequently, constructs the corresponding TDI solutions.

Specifically, the communicator in question can be obtained based on Eq.~\eqref{sys1}, where one uses the subscript $x$ to denote the first term and $y$ for the second term of the commutator.
The communicator consists of two terms of the same size $[D_{x_{1} x_{2} \dots x_{n}}, D_{y_{1} y_{2} \dots y_{n}}]$, and it satisfies the following relation~\cite{D2009}
\begin{equation}\label{sys1}
	\begin{gathered}
		{\left[D_{x_{1} x_{2} \ldots x_{n}}, D_{y_{1} y_{2} \ldots y_{n}}\right] \phi(t)} 
		=\left(\sum_{i=1}^{n} L_{x_i} \sum_{j=1}^{n} \dot{L}_{y_j}-\sum_{j=1}^{n} L_{y_j} \sum_{i=1}^{n} \dot{L}_{x_i}\right) 
		\dot{\phi}\left(t-\sum_{k=1}^{n} L_{x_{k}}-\sum_{k'=1}^{n} L_{y_{k'}}\right) ,
	\end{gathered}
\end{equation}
where a dot indicates the time derivative, and $\phi(t)$ is an arbitrary function of time.
A proof of Eq.~\eqref{sys1} is given in~\cite{cTDI2gen}, which further generalizes to take into account time-advance operators $D_{\bar{i}, (\bar{i'})}$, so that the subscripts $x, y = i, i', \bar{i}, \bar{i}'$. 
It is not difficult to show that the first factor on the r.h.s. of Eq.~\eqref{sys1} vanishes as long as
\begin{equation}\label{permuXY}
	y_i=x_{\pi(i)} .
\end{equation}
where $\pi\in \mathcal{S}_n$ is an element of the permutation group of degree $n$. 
In other words, Eq.~\eqref{sys1} vanishes when $\{y_{1} y_{2} \ldots y_{n}\}$ is an arbitrary permutation of $\{x_{1} x_{2} \ldots x_{n}\}$.
Moreover, as elaborated in~\cite{cTDI2gen}, the difference between two monomials in $a$ and $b$ can be written as summation of multiplier of $(1-a)$ or $(1-b)$.
As a result, the corresponding coefficients furnish a solution for $\alpha$ and $\beta$ of Eq.~\eqref{nMTDIgen}, giving rise to a feasible TDI solution.

As an example, the Michelson-X combination can be derived by the commutator
\begin{equation}\label{M16V2}
\Delta = [ba, ab] ,
\end{equation} 
whose TDI coefficients reads
\begin{equation}\label{M16}
	\begin{aligned}
	q_{1} &=\alpha= 1-b-ba+ab^2,\\
	q_{1^{'}} &=\beta= - (1-a-ab+ba^2) .
	\end{aligned}
\end{equation}

\section{The second-order combinatorial algebraic approach}\label{sec3}

In this section, we first propose two novel classes of second-generation TDI solutions by propositions~\ref{prop1},~\ref{prop2}, and~\ref{prop3}.
In analogy to the algebraic approach discussed in the previous section, these solutions are also based on a particular form of commutator that vanishes.
Unlike Eq.~\eqref{sys1}, the proposed formalisms are based on second-order commutators. 
Furthermore, we show that the solution space can be further expanded using corrollaries~\ref{coro1} and~\ref{coro2}.

\begin{prop}\label{prop1}
The following second-order commutator vanishes when the contributions associated with the second and higher-order time derivatives are ignored
\begin{equation}\label{mi1}
\Delta_1 = \left[C, [A, B]\right] ,
\end{equation}
where $A =\sum_i a_i$, $B =\sum_j b_j$, and $C =\sum_k c_k$ are arbitrary polynomials, and $a_i, b_j, c_k$ are monomials, in the time-delay and time-advance operators.
\end{prop}

\begin{prop}\label{prop2}
The following product of first-order commutators vanishes when the contributions associated with the second and higher-order time derivatives are ignored
\begin{equation}\label{mi2}
\Delta_2 = [C, D][A, B] ,
\end{equation}
where $A, B, C, D$ are polynomials.
\end{prop}

More details regarding the proofs of propositions~\ref{prop1} and~\ref{prop2} are relegated to Appendix~\ref{appA}.

\begin{prop}\label{prop3}
A commutator constructed by employing either proposition 1 or 2 can always be written as a summation of multipliers of $(1-a)$ or $(1-b)$.
Subsequently, the TDI coefficients can be derived.
\end{prop}

A general process for proposition 3 is given in Appendix~\ref{appB}, closely following the discussions in~\cite{cTDI2gen}.

In what follows, we elaborate on a few examples based on the above propositions.
Regarding proposition 1, a Michelson-type solution for Eq.~\eqref{nMTDIgen} reads
\begin{equation}\label{Mi-mi1}
\Delta_1=[a, [a, b]]=2aba - a^2b -ab^2 .
\end{equation}
By noticing
\begin{equation}\label{M1786}
\begin{aligned}
aba &= -ab(1-a) - a(1-b) -(1-a) +1 ,\\
a^2b &= -a^2 (1-b) - a(1-a) -(1-a) +1 ,\\
ba^2 &= -ba (1-a) -b (1-a) - (1-b) +1 .
\end{aligned}
\end{equation}
Eq.~\eqref{Mi-mi1} implies the following TDI coefficients
\begin{equation}\label{Mi-mi1-cos}
	\begin{aligned}
	q_{1} &=\alpha= -2ab -2 + a + 1 +ba +b = -2ab+ba + a + b -1,\\
	q_{1^{'}} &=\beta= -2a + a^2 +1 = a^2 - 2a +1 .
	\end{aligned}
\end{equation}

When compared against Michelson-X solution Eq.~\eqref{M16V2}, the solution Eq.~\eqref{Mi-mi1} involves three terms instead of two.
Different from Eq.~\eqref{sys1}, the two terms of the commutator are not necessarily of equal length nor an equal number of factors, such as $[a, [a, ab]]$.
Moreover, in its lowest order, Eq.~\eqref{Mi-mi1} is featured by a compact size.
To our knowledge, the solution Eq.~\eqref{Mi-mi1} lies beyond the solution space explored in most literature, inclusively the preceding combinatorial algebraic algorithms.

An example of proposition 2 is
\begin{equation}\label{Mi-mi2}
\Delta_2=[a, b][a, b]=abab+baba-ab^2a-ba^2b .
\end{equation}
By noticing
\begin{equation}\label{M1787}
\begin{aligned}
abab &= -aba(1-b) - ab(1-a) -a(1-b) -(1-a) +1 ,\\
baba &= -bab(1-a) - ba(1-b) -b(1-a) -(1-b)+1 ,\\
ab^2a &= -ab^2 (1-a) - ab(1-b) -a(1-b) - (1-a) +1 ,\\
ba^2b &= -ba^2 (1-b) -ba(1-a) -b(1-a) - (1-b) +1 .
\end{aligned}
\end{equation}
Eq.~\eqref{Mi-mi1} implies the following TDI coefficients
\begin{equation}\label{Mi-mi2-cos}
	\begin{aligned}
	q_{1} &=\alpha= -ab -1 -bab -b +ab^2 +1 +ab +b= ab^2-bab-ab+ba  ,\\
	q_{1^{'}} &=\beta= -aba -a -ba -1+ab +a +ba^2 +1= -aba+ba^2+ab-ba   .
	\end{aligned}
\end{equation}

Compared with the Michelson-X solution Eq.~\eqref{M16V2}, the solution Eq.~\eqref{Mi-mi2} double the number of terms of equal size.
Again, the terms involved in the commutator are not necessarily of equal length. 
Therefore the solution space is more significant than the first-order commutator approach.
In its lowest order, the solution Eq.~\eqref{Mi-mi2} is of equal size compared to the lowest-order second-generation geometric TDI ones.
However, when one goes to a higher order, it may give rise to a trajectory featuring an odd count of total links, such as $[a, b][a, b^2]$, which is not included in Eq.~\eqref{sys1} by definition.
Moreover, the following corollaries expand the solution space while extending to the higher-order cases.

\begin{coro}\label{coro1}
A higher-order commutator can be constructed by replacing the commutator $[A, B]$ in propositions 1 and 2 with a lower-order one formed using the propositions, namely,
\begin{equation}\label{mi3}
\Delta_3 = \left[C, \Delta_1\right] ,
\end{equation}
or
\begin{equation}\label{mi4}
{\Delta_3}' = \left[C, D\right]\Delta_2 .
\end{equation}
The resulting commutator readily vanishes when the contributions associated with the second and higher-order time derivatives are ignored.
\end{coro}

\begin{coro}\label{coro2}
A commutator constructed by the linear combination of the commutators derived using propositions 1 - 2 and corollary 1 vanishes when the contributions associated with the second and higher-order time derivatives are ignored.
Here, linear combination coefficients are defined as arbitrary polynomials of time-translation operators, which can be multiplied either to the left or right of the existing commutators.
\end{coro}

An example of corollary~\ref{coro2} is
\begin{equation}\label{Mi-mi4}
\Delta_4=\Delta_2 - b\Delta_1=abab+baba-ab^2a-ba^2b - b\left(2aba - a^2b -ab^2\right) = abab-baba-ab^2a+bab^2 .
\end{equation}
By noticing Eqs.~\eqref{M1787} and
\begin{equation}\label{M1787b}
\begin{aligned}
bab^2 &= -bab (1-b) -ba(1-b) -b(1-a) - (1-b) +1 .
\end{aligned}
\end{equation}
Eq.~\eqref{Mi-mi4} implies the following TDI coefficients
\begin{equation}\label{Mi-mi4-cos}
	\begin{aligned}
	q_{1} &=\alpha= -ab -1 +bab +b +ab^2 +1 -ba -b= ab^2+bab-ab-ba  ,\\
	q_{1^{'}} &=\beta= -aba -a +ba +1+ab +a -bab -1= -aba-bab+ab+ba  .
	\end{aligned}
\end{equation}

\section{Applications to different types of second-order TDI combinations}\label{sec4}

In this section, we apply the proposed algorithm to TDI solutions of different types, such as the Monitor, Beacon, Relay, Sagnac, and full symmetric Sagnac ones.
Besides, we elaborate on the corresponding sensitivity curves compared with the standard second-generation solutions.
The specific forms of the response functions and residual noise power spectral densities are relegated to Appendix~\ref{appC}.
We will only focus on the two lowest-order solutions given by Eqs.~\eqref{Mi-mi1-cos} and~\eqref{Mi-mi2-cos}.
It is noted that higher-order solutions can be similarly constructed by employing the algorithm elaborated in the last section.

\subsubsection{Michelson-type solutions}

For the Michelson-type solutions, Eq.~\eqref{Mi-mi1-cos} gives
\begin{equation}
	\begin{aligned}
		q_{1} &= \alpha = - 2D_{33'2'2} + D_{2'233'} + D_{33'} + D_{2'2} - 1,\\
		q_{1^{'}} &= \beta =D_{33'33'}- 2D_{33'} + 1,
	\end{aligned}
\end{equation}
which gives rise to
\begin{align}\label{XDelta1}
X_{\Delta_1} =& ( - 2D_{33'2'2}+ D_{2'233'} + D_{33'} + D_{2'2} - 1){\eta _1} + ( {{D_{33'33'2'}} - 2{D_{33'2'}} + {D_{2'}}}){\eta _3}\notag\\
 +& ({D_{33'33'} - 2{D_{33'}} + 1}){\eta _{1'}} +(- 2{D_{33'2'23}} + {D_{2'233'3}} + {D_{33'3}} + {D_{2'23}} - {D_3}){\eta _{2'}} .
\end{align}
Similarly, Eq.~\eqref{Mi-mi2-cos} gives
\begin{equation}
	\begin{aligned}
		q_{1} &= \alpha =D_{33'2'22'2} + D_{2'233'} - D_{33'2'2}- D_{2'233'2'2},\\
		q_{1^{'}} &= \beta = D_{2'233'33'} + D_{33'2'2} -D_{2'233'} - D_{33'2'233'},
	\end{aligned}
\end{equation}
which leads to
\begin{align}\label{XDelta2}
X_{\Delta_2}=& \left( {{D_{33'2'22'2}} + {D_{2'233'}} - {D_{33'2'2}} - {D_{2'233'2'2}}} \right){\eta _1} + \left( {{D_{2'233'33'2'}} + {D_{33'2'22'}} - {D_{2'233'2'}} - {D_{33'2'233'2'}}} \right){\eta _3}\notag\\
 +& \left( {{D_{2'233'33'}} + {D_{33'2'2}} - {D_{2'233'}} - {D_{33'2'233'}}} \right){\eta _{1'}} + \left( {{D_{33'2'22'23}} + {D_{2'233'3}} - {D_{33'2'23}} - {D_{2'233'2'23}}} \right){\eta _{2'}}.
\end{align}
Lastly, Eq.~~\eqref{Mi-mi4-cos} gives
\begin{equation}
	\begin{aligned}
{q_1} =&  - {D_{33'2'2}} + {D_{33'2'22'2}} + {D_{2'233'2'2}} + {D_{2'2}} - {D_{2'22'233'}} - {D_{2'22'2}},\\
{q_{1'}} =&  - {D_{33'2'233'}} + {D_{33'2'2}} + {D_{2'233'}} - {D_{2'2}}
	\end{aligned}
\end{equation}
which leads to
\begin{align}\label{XDelta4}
X_{\Delta _4}=& \left( { - {D_{33'2'2}} + {D_{33'2'22'2}} + {D_{2'233'2'2}} + {D_{2'2}} - {D_{2'22'233'}} - {D_{2'22'2}}} \right){\eta _{\rm{1}}}\notag\\
 +& \left( { - {D_{33'2'233'2'}} + {D_{33'2'22'}} + {D_{2'233'2'}} - {D_{2'22'}}} \right){\eta _{\rm{3}}}\notag\\
+& \left( { - {D_{33'2'233'}} + {D_{33'2'2}} + {D_{2'233'}} - {D_{2'2}}} \right){\eta _{{\rm{1'}}}}\notag\\
 +& \left( { - {D_{33'2'23}} + {D_{33'2'22'23}} + {D_{2'233'2'23}} + {D_{2'23}} - {D_{2'22'233'3}} - {D_{2'22'23}}} \right){\eta _{{\rm{2'}}}}.
\end{align}

\begin{figure}[h]
\includegraphics[width=0.6\textwidth]{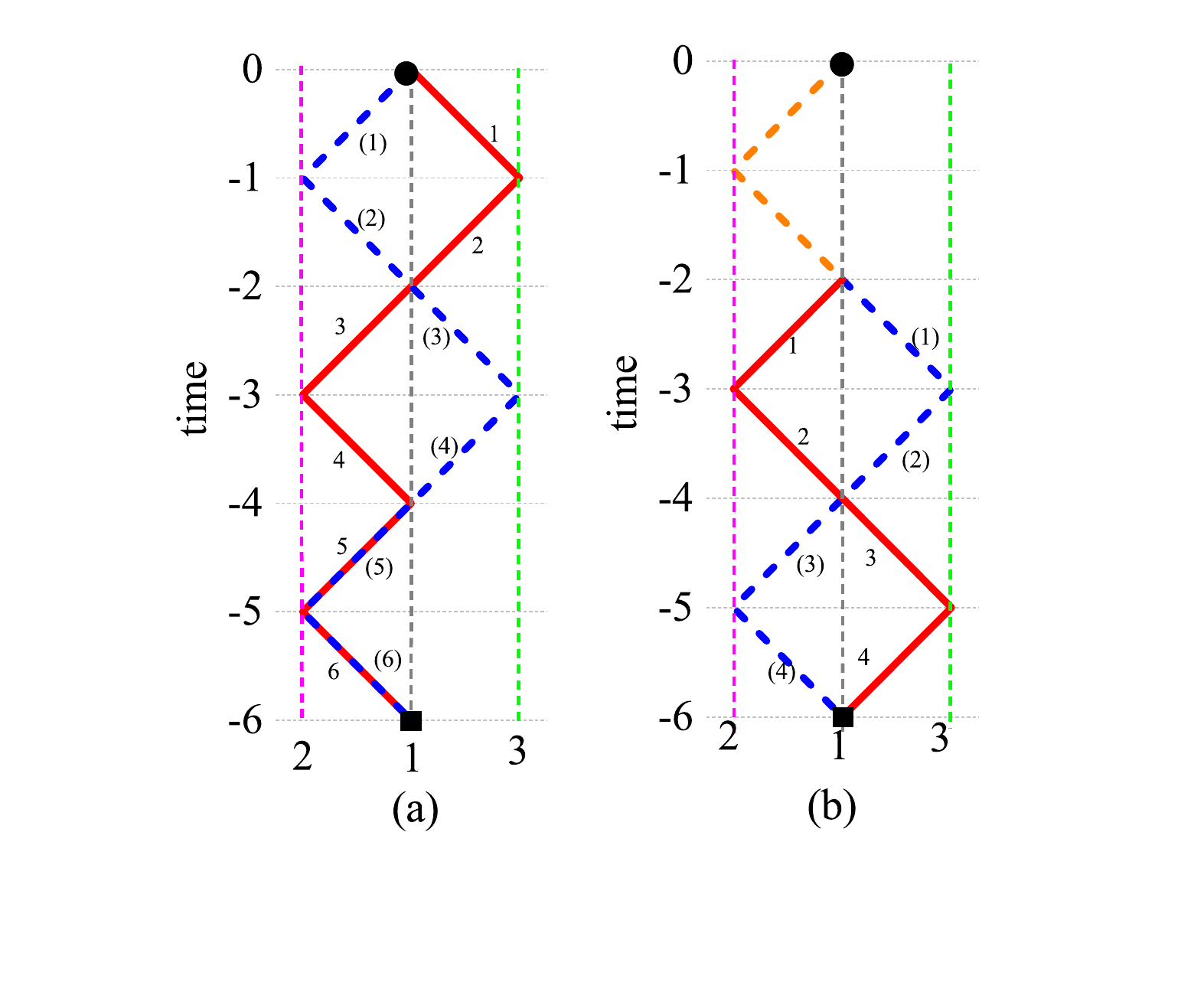}\ \ \includegraphics[width=0.3\textwidth]{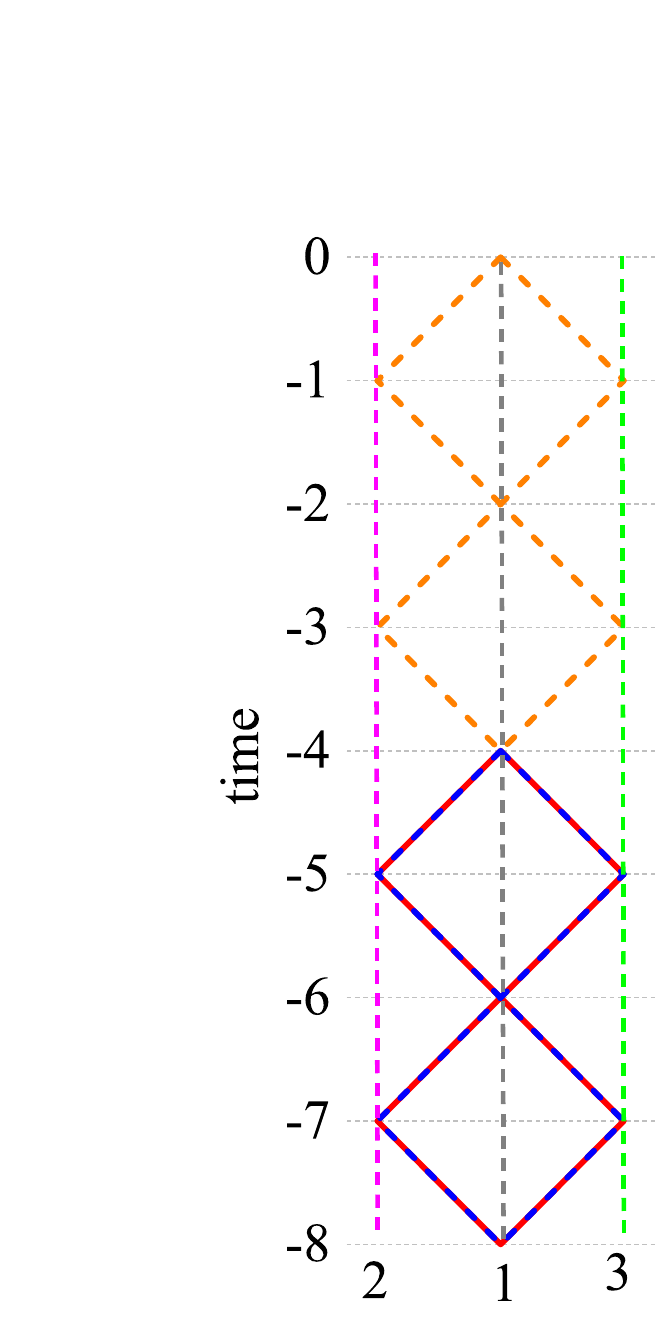}\\
\includegraphics[width=0.6\textwidth]{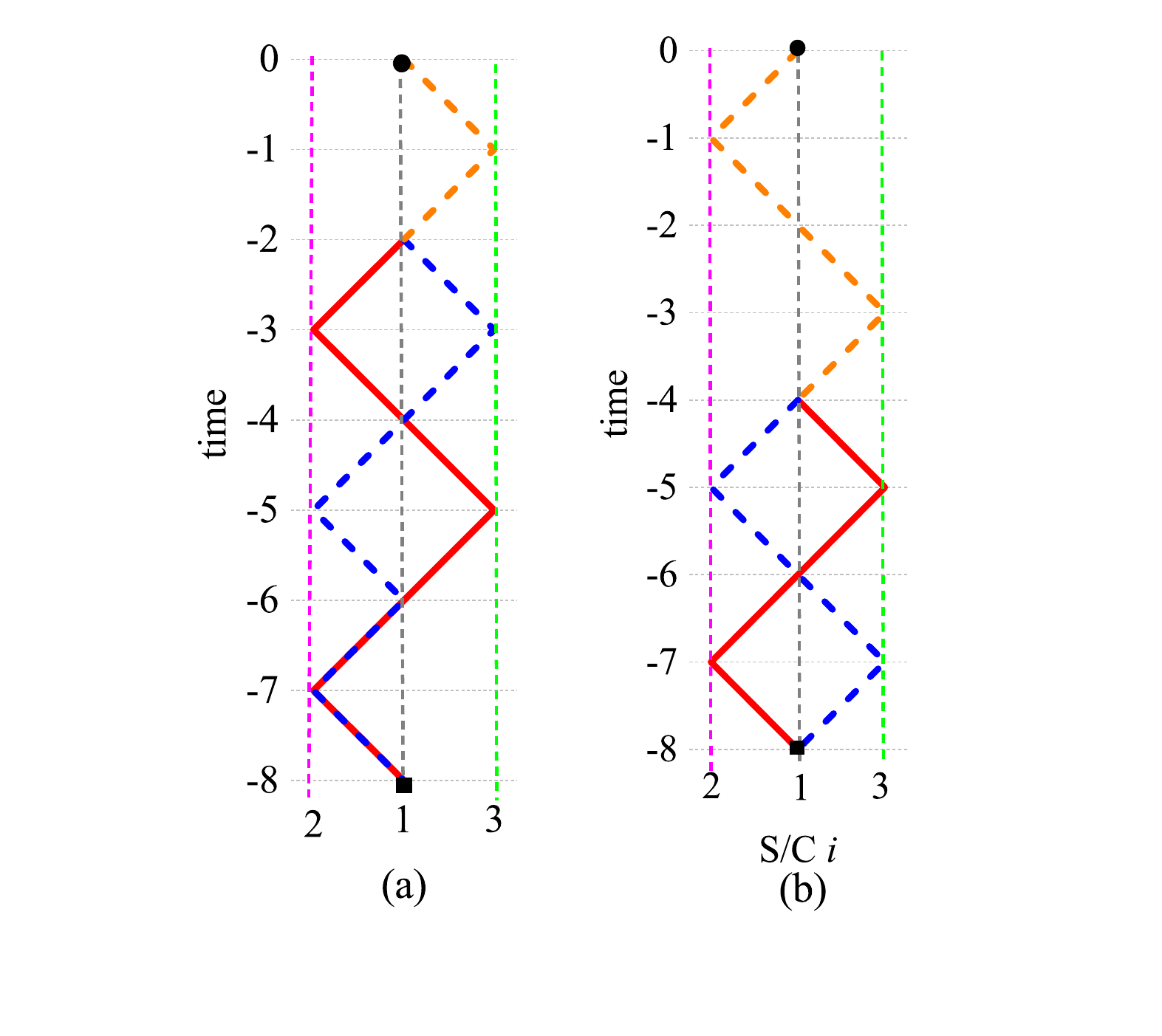}
\caption{\label{fig2} The space-time diagrams for $X_{\Delta_1}$ (top left), $X_{\Delta_2}$ (bottom), and $X_{\Delta_4}$ (top right).
The solid red and dashed blue line segments correspond to the two different light routes, while the dashed orange line segments indicate additional time-displacement operations common to the data streams in question.}
\end{figure}

Now we proceed to discuss the relationship between the above solutions and those obtained by the standard geometric TDI approach.
It is not difficult to verify that all three solutions can be obtained by the geometric TDI approach for a given link number using the method of exhaustion~\cite{Geo-sister}.
One might have expected that solution Eq.~\eqref{XDelta1} belongs to the subset of ten-link geometric TDI combinations by simply counting the subscript indices of individual terms.
However, as it turns out, it is a member of the twenty-link family.
This is because the degree of a geodesic TDI solution is not necessarily governed by the term which contains the most subscript indices, as one might have to successively multiply the appropriate inverse operators from the left in order to retrieve the terms which correspond to the links suffering less time displacement operations.
In other words, for a valid geometric TDI solution, the terminal time instant is not necessarily at $t=0$ as long as the optical path encloses itself.
On the other hand, it is observed that the degree of a geometric TDI solution must equate to the total number of terms.
To be specific, the corresponding light propagation trajectories are shown in the space-time diagrams in Fig.~\ref{fig2}.

To be more explicit, for $X_{\Delta_1}$, its geometric TDI counterpart can be seen more transparently by rewriting Eq.~\eqref{XDelta1} as
\begin{align}\label{geoXDelta1}
X_{\Delta_1} =& - \left( {{\eta _{\rm{1}}}{\rm{ + }}{D_3}{\eta _{{\rm{2'}}}}{\rm{ + }}{D_{33'}}{\eta _{{\rm{1'}}}}{\rm{ + }}{D_{33'2'}}{\eta _{\rm{3}}}{\rm{ + }}{D_{33'2'2}}{\eta _{\rm{1}}}{\rm{ + }}{D_{33'2'23}}{\eta _{{\rm{2'}}}}} \right)\notag\\
+& \left( {{\eta _{{\rm{1'}}}}{\rm{ + }}{D_{2'}}{\eta _{\rm{3}}}{\rm{ + }}{D_{2'2}}{\eta _{\rm{1}}}{\rm{ + }}{D_{2'23}}{\eta _{{\rm{2'}}}}{\rm{ + }}{D_{2'233'}}{\eta _{\rm{1}}}{\rm{ + }}{D_{2'233'3}}{\eta _{{\rm{2'}}}}} \right)\notag\\
 -& \left( {{D_{33'}}{\eta _{{\rm{1'}}}}{\rm{ + }}{D_{33'2'}}{\eta _{\rm{3}}}{\rm{ + }}{D_{33'2'2}}{\eta _{\rm{1}}}{\rm{ + }}{D_{33'2'23}}{\eta _{{\rm{2'}}}}} \right)\notag\\
+ &\left( {{D_{33'}}{\eta _{\rm{1}}}{\rm{ + }}{D_{33'3}}{\eta _{{\rm{2'}}}}{\rm{ + }}{D_{33'33'}}{\eta _{{\rm{1'}}}}{\rm{ + }}{D_{33'33'2'}}{\eta _{\rm{3}}}} \right).
\end{align}
The corresponding light propagation trajectories are shown in the space-time diagrams in the top-left plot of Fig.~\ref{fig2}.
We note that the solution can be further decomposed into two first-generation combinations, corresponding to the first two and last two lines of Eq.~\eqref{geoXDelta1}.
In particular, the data stream associated with the first line of Eq.~\eqref{geoXDelta1} is indicated by the dashed blue line segments denoted by ``(1)''-``(6)'' shown in left subplot (a).
The data stream associated with the second line of the equation is represented by the solid red line segments denoted by ``1''-``6'' also shown in the left subplot.
It is apparent that these two six-link trajectories form a closed trajectory.
Similarly, the data stream associated with the third line of Eq.~\eqref{geoXDelta1} is indicated by the dashed blue line segments denoted by ``(1)''-``(4)'' shown in the right subplot (b), where the dashed orange line segments correspond to additional time-displacements common to the data streams in question.
The data stream associated with the fourth line of the equation is given by the solid red line segments denoted by ``1''-``4'' also shown in the right subplot.
Again, it is readily observed that the last two lines form a second closed trajectory. 
%Moreover, among the two close trajectories shown in the top-left plot of Fig.~\ref{fig2}, the left one furnishes a valid twelve-link geometric TDI solution.
Similarly, the space-time diagrams of the geometric TDI counterparts of the solutions $X_{\Delta_2}$ and $X_{\Delta_4}$ are also shown in Fig.~\ref{fig2}.

The corresponding sensitivity curves are evaluated below in Appendix~\ref{appC} and shown in Fig.~\ref{fig3} compared with the sixteen-link Michelson-X solution Eq.~\eqref{M16}.
It is observed that the sensitivity performance of the solution $X_{\Delta_1}$ is analytically identical to the Michelson-X one $X_1^{16}$~\cite{TDI2-1}.

\begin{figure}[h]
\includegraphics[width=0.45\textwidth]{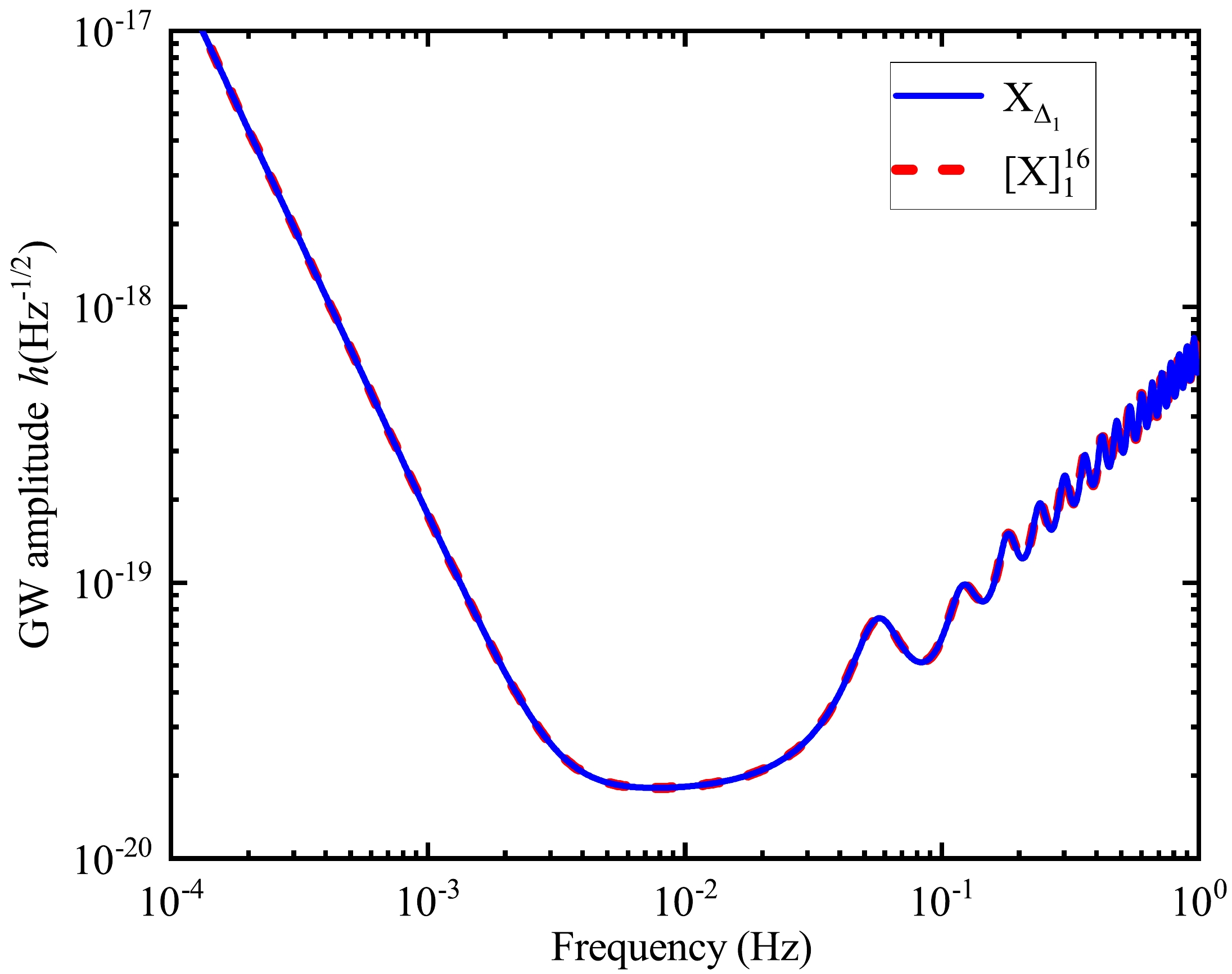}\ \ \includegraphics[width=0.45\textwidth]{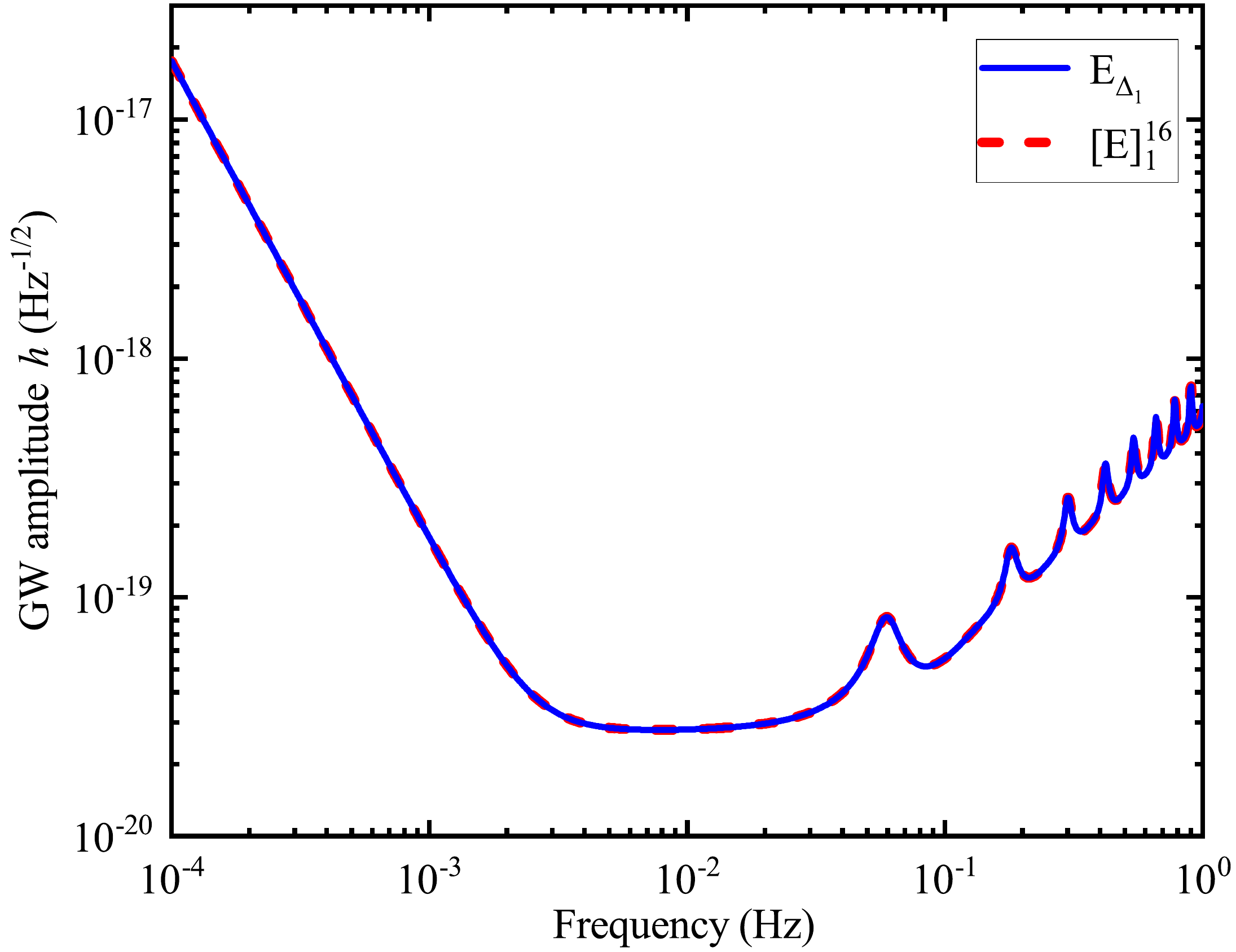}\\
\includegraphics[width=0.45\textwidth]{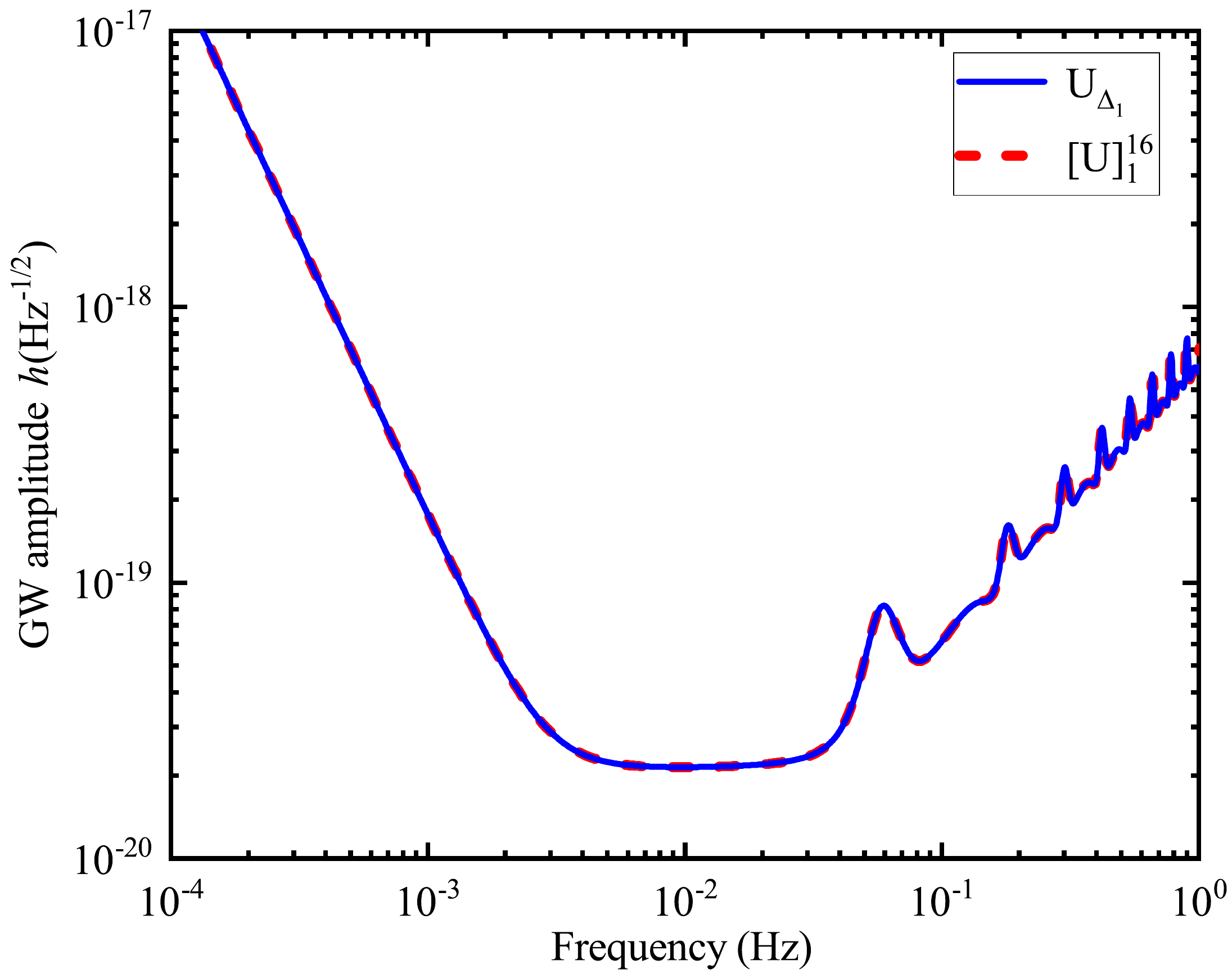}\ \ \includegraphics[width=0.45\textwidth]{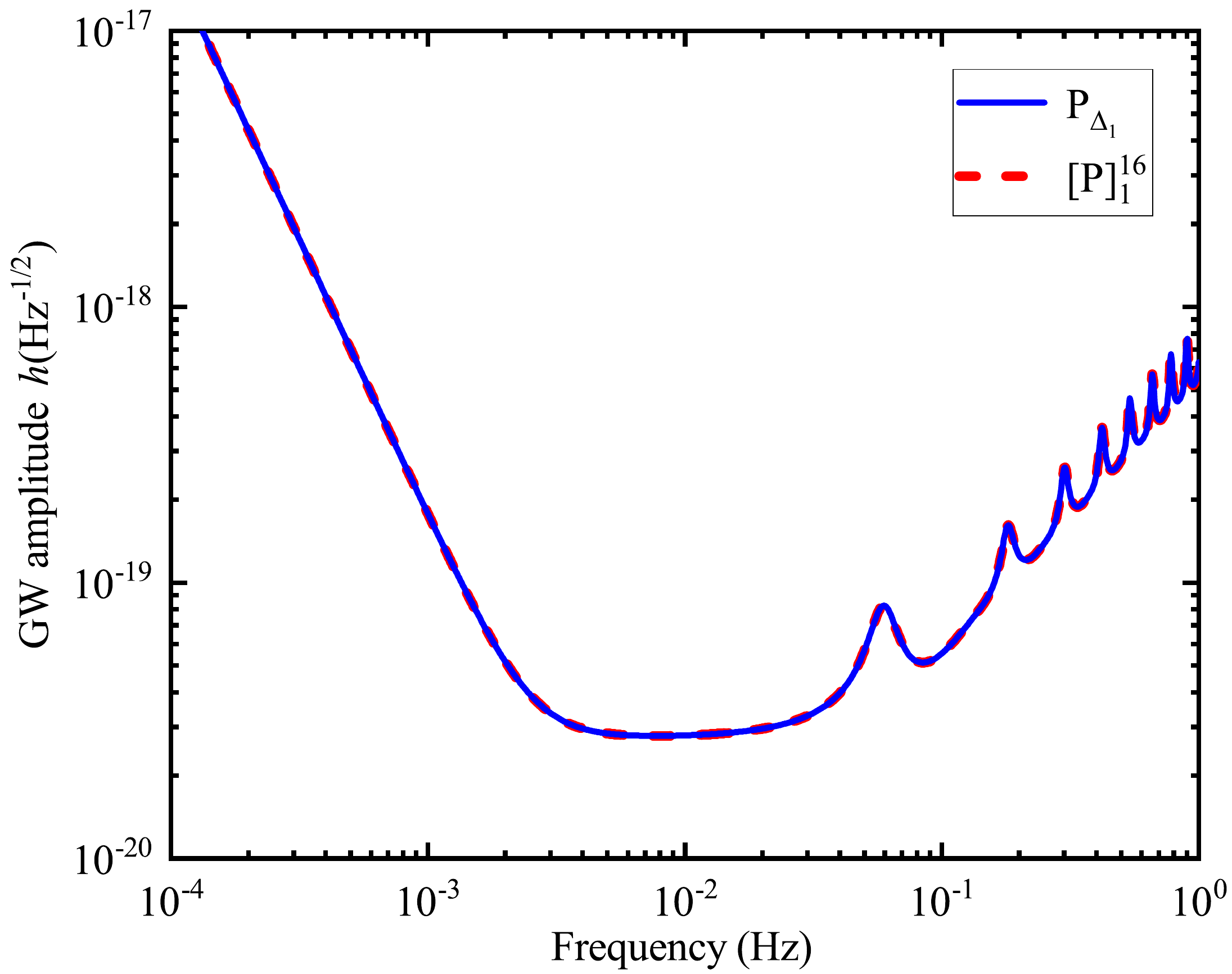}\\
\includegraphics[width=0.45\textwidth]{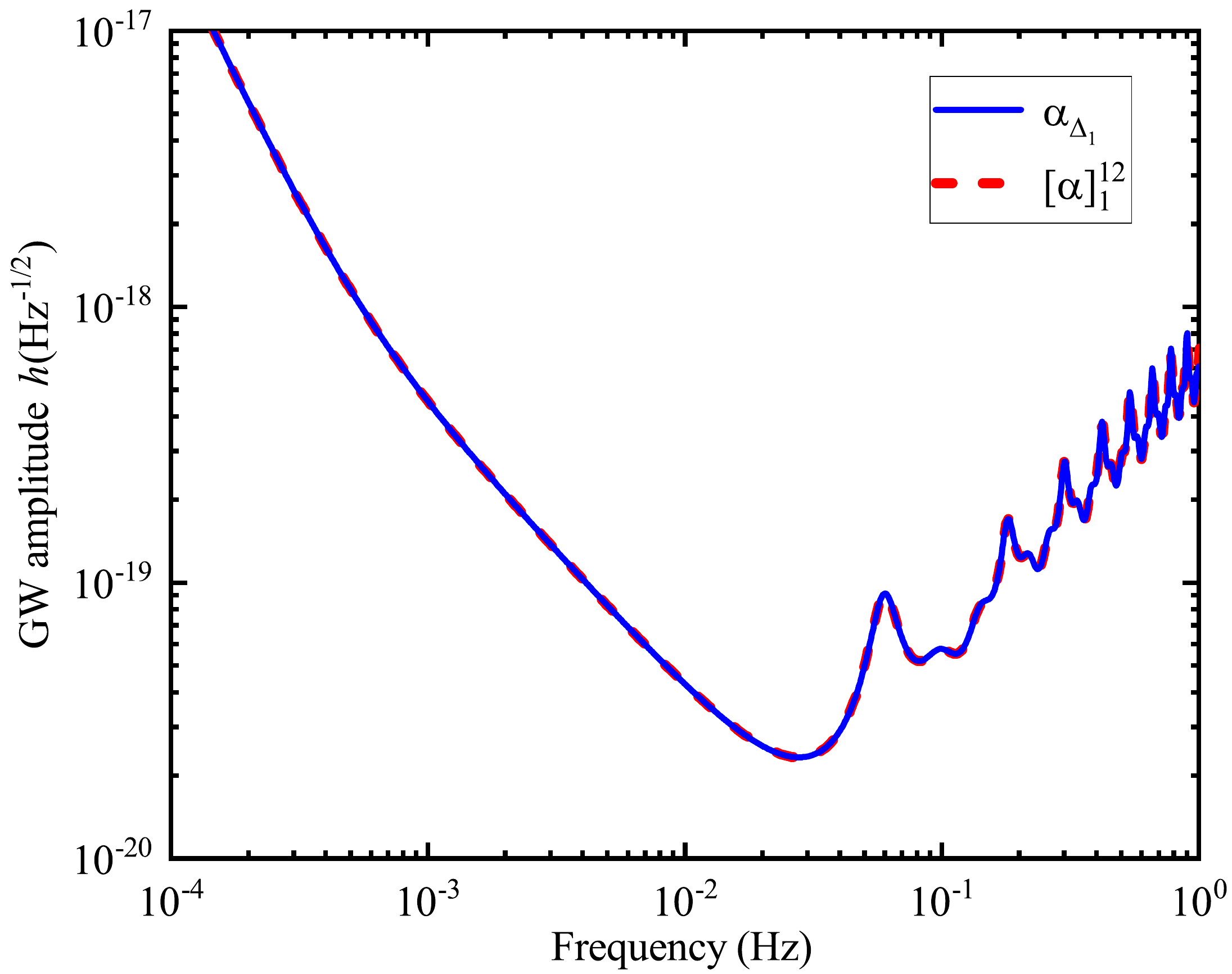}\ \ \includegraphics[width=0.45\textwidth]{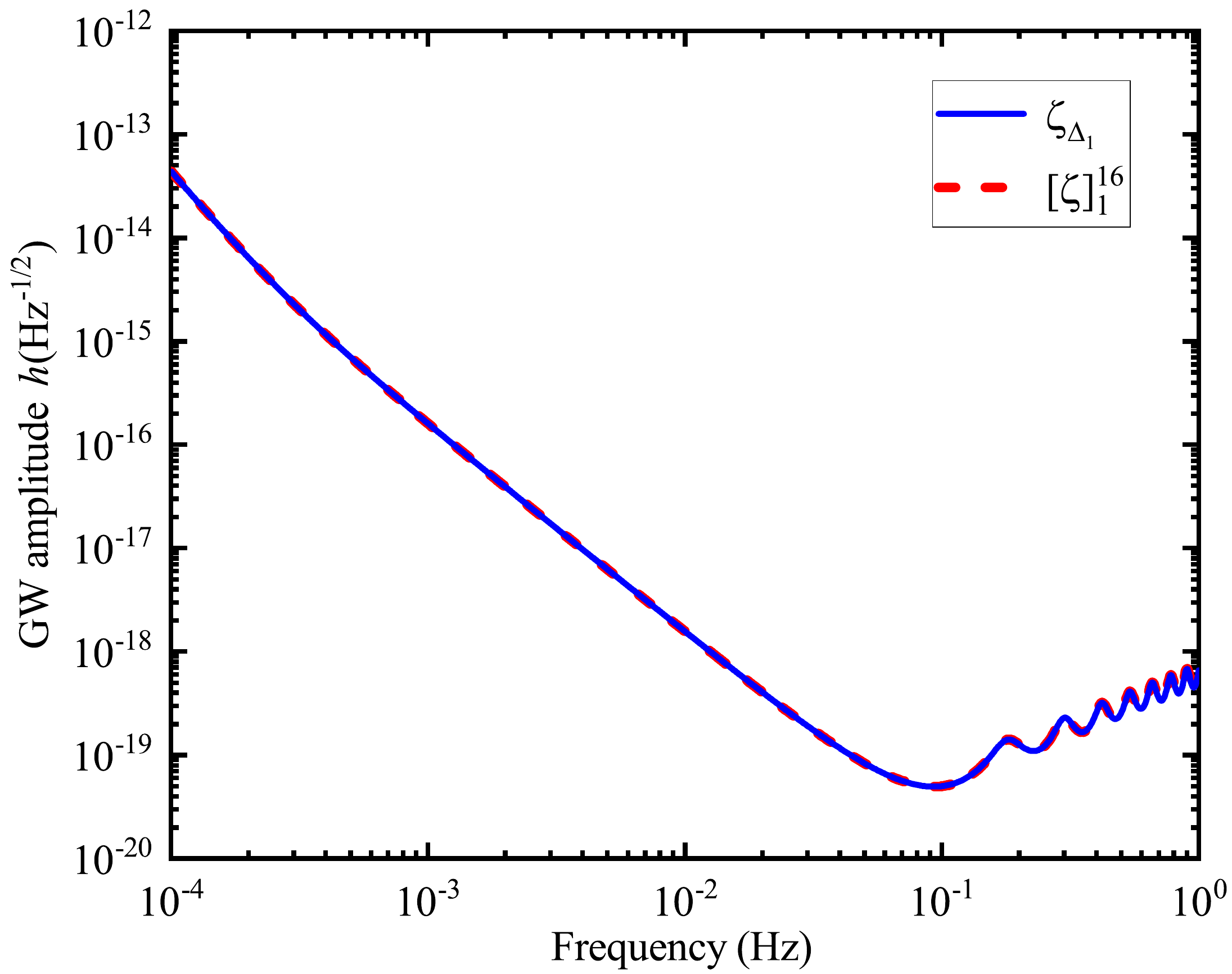}\\
\caption{\label{fig3} (Color online) The sensitivity curves for Michelson-X (top left), Monitor-E (top right), Beacon (middle left), Relay (middle right), Sagnac (bottom left), and fully-symmetric Sagnac (bottom right) type combinations obtained using the proposed algorithm.
The solutions derived in this study are presented in solid blue curves, while the geometric TDI ones are shown in dashed red curves.}
\end{figure}

\subsubsection{Monitor-type solutions}

Following the arguments given in~\cite{cTDI2gen}, one considers the following two constraint equations for the second-generation Monitor-E combinations
\begin{equation}\label{Moncons}
	q_{3}=0, q_{2^{'}}=0.
\end{equation}
By substituting Eq.~\eqref{Moncons} into Eq.~\eqref{TDIeq}, and eliminating $q_1$ and $q_{1'}$, one finds
\begin{equation}\label{MonitorTDI}
	q_{2}(D_{\bar{3}}-D_{1\bar{2}^{'}})+q_{3^{'}}(D_{\bar{2}^{'}}-D_{1^{'}\bar{3}})=0 .
\end{equation}
The above equation is essentially identical to Eq.~\eqref{nMTDIgen} by recognizing $\alpha=q_{2}D_{\bar{3}}, \beta=q_{3^{'}}D_{\bar{2}^{'}}$, $a=D_{31\bar{2}^{'}}$, and $b=D_{2^{'}1^{'}\bar{3}}$.

For the commutator $[a, [a, b]]$, the corresponding TDI solution expressed in terms of the coefficients $q_2, q_{3^{'}}$ reads
\begin{equation}
	\begin{aligned}
		q_{2} &= \alpha{D_3} = - 2{D_{311'}} + {D_{2'1'1\bar 2'{\rm{3}}}} + {D_{31\bar 2'{\rm{3}}}} + {D_{2'1'}} - {D_3},\\
		q_{3^{'}} &= \beta {D_{2'}} = {D_{31\bar 2'31}} - 2{D_{31}} + {D_{2'}},
	\end{aligned}
\end{equation}
which gives rise to
\begin{align}
E_{\Delta_1}=& \left( {{D_{2'1'1\bar 2'}} + {D_{31\bar 2'}} - {D_{31\bar 2'31{\rm{1'\bar 3}}}} - {\rm{1}}} \right){\eta _1} + \left( { - 2{D_{311'}} + {D_{2'1'1\bar 2'{\rm{3}}}} + {D_{31\bar 2'{\rm{3}}}} + {D_{2'1'}} - {D_3}} \right){\eta _2}\notag\\
 +&\left( { - {D_{31\bar 2'}} + 1 + 2{D_{311'1\bar 2'}} - {D_{2'1'1\bar 2'{\rm{31}}\bar 2'}} - {D_{2'1'1\bar 2'}}} \right){\eta _{1'}} + \left( {{D_{31\bar 2'31}} - 2{D_{31}} + {D_{2'}}} \right){\eta _{3'}} .
\end{align}

Also, the solution corresponds to $[a, b][a, b]$ reads
\begin{equation}
	\begin{aligned}
		q_{2} &= \alpha{D_3}  = {D_{311'\bar 32'1'}} + {D_{2'1'1\bar 2'3}} - {D_{311'}} - {D_{2'1'11'}},\\
		q_{3^{'}} &= \beta{D_{2'}} ={D_{2'1'1\bar 2'31}} + {D_{311'\bar 32'}} - {D_{2'1'1}} - {D_{311'1}},
	\end{aligned}
\end{equation}
which leads to
\begin{align}
E_{\Delta_2}=& \left( {{D_{2'1'1\bar 2'}} - {D_{311'\bar 3}} - {D_{2'1'1\bar 2'311'\bar 3}} + {D_{311'11'\bar 3}}} \right){\eta _1} + \left( {{D_{311'\bar 32'1'}} + {D_{2'1'1\bar 2'3}} - {D_{311'}} - {D_{2'1'11'}}} \right){\eta _2}\notag\\
 +& \left( {{D_{311'\bar 3}} - {D_{2'1'1\bar 2'}} - {D_{311'\bar 32'1'1\bar 2'}} + {D_{2'1'11'1\bar 2'}}} \right){\eta _{1'}} + \left( {{D_{2'1'1\bar 2'31}} + {D_{311'\bar 32'}} - {D_{2'1'1}} - {D_{311'1}}} \right){\eta _{3'}}.
\end{align}

We show in Fig.~\ref{fig3} the sensitivity curve which is found to be identical to the standard second-generation Monitor-E combination $E_1^{16}$~\cite{TDI2-1}.

\subsubsection{Relay-type solutions}

For the Relay type, we consider the following constraints
\begin{equation}\label{relay}
	q_{2}=0, q_{3}=0.
\end{equation}
which gives
\begin{equation}
	q_{2^{'}}(D_{\bar{3}}-D_{3^{'}})+q_{3^{'}}(D_{\bar{2}^{'}}-D_{1^{'}\bar{3}})=0.
\end{equation}
Again, it is equivalent to Eq.~\eqref{nMTDIgen} by identifying $\alpha=q_{2^{'}}D_{\bar{3}}$, $\beta=q_{3^{'}}D_{\bar{2}^{'}}$, $a=D_{33^{'}}$ and $b=D_{2^{'}1^{'}\bar{3}}$.

For the commutator $[a, [a, b]]$, the corresponding TDI solution expressed in terms of the coefficients $q_{2'}, q_{3^{'}}$ reads
\begin{equation}
	\begin{aligned}
		q_{2'} &= \alpha D_{3}=- 2{D_{33'2'1'}} + {D_{2'1'3'3}} + {D_{33'3}} + {D_{2'1'}} - {D_3},\\
		q_{3^{'}} &= \beta D_{2'}= {D_{33'33'2'}} - 2{D_{33'2'}} + {D_{2'}},
	\end{aligned}
\end{equation}
which gives rise to
\begin{align}
U_{\Delta_1}=& \left( {{D_{2'1'3'}} + {D_{33'}} - 1 - {D_{33'33'2'1'\bar 3}}} \right){\eta _1} + \left( {{D_{33'33'}} - 2{D_{33'}} + 1} \right){\eta _{1'}}\notag\\
 + &\left( { - 2{D_{33'2'1'}} + {D_{2'1'3'3}} + {D_{33'3}} + {D_{2'1'}} - {D_3}} \right){\eta _{2'}} + \left( {{D_{33'33'2'}} - 2{D_{33'2'}} + {D_{2'}}} \right){\eta _{3'}} .
\end{align}

Also, the solution corresponding to the commutator $[a, b][a, b]$ reads
\begin{equation}
	\begin{aligned}
		q_{2'} &= \alpha D_{3}= {D_{33'2'1'\bar 32'1'}} + {D_{2'1'3'{\rm{3}}}} - {D_{33'2'1'}} - {D_{2'1'3'2'1'}},\\
		q_{3^{'}} &= \beta D_{2'}= {D_{2'1'3'33'{\rm{2'}}}} + {D_{33'2'1'\bar 3{\rm{2'}}}} - {D_{2'1'3'{\rm{2'}}}} - {D_{33'2'1'3'{\rm{2'}}}},
	\end{aligned}
\end{equation}
which leads to
\begin{align}
U_{\Delta_2}=& \left( {{D_{2'1'3'}} - {D_{33'2'1'\bar 3}} - {D_{2'1'3'33'{\rm{2'1'}}\bar 3}} + {D_{33'2'1'3'{\rm{2'1'}}\bar 3}}} \right){\eta _1} + \left( {{D_{2'1'3'33'}} + {D_{33'2'1'\bar 3}} - {D_{2'1'3'}} - {D_{33'2'1'3'}}} \right){\eta _{1'}}\notag\\
 +& \left( {{D_{33'2'1'\bar 32'1'}} + {D_{2'1'3'{\rm{3}}}} - {D_{33'2'1'}} - {D_{2'1'3'2'1'}}} \right){\eta _{2'}} + \left( {{D_{2'1'3'33'{\rm{2'}}}} + {D_{33'2'1'\bar 3{\rm{2'}}}} - {D_{2'1'3'{\rm{2'}}}} - {D_{33'2'1'3'{\rm{2'}}}}} \right){\eta _{3'}}.
\end{align}
We show in Fig.~\ref{fig3} the resulting sensitivity curve, which is identical to that of the standard second-generation Relay combination $U_1^{16}$~\cite{TDI2-1}.

\subsubsection{Beacon-type solutions}

For the Beacon type, we consider the following constraints
\begin{equation}
	q_{3}=0, q_{3^{'}}=0,
\end{equation}
which gives
\begin{equation}
	q_{1}(1-D_{33^{'}})+q_{1^{'}}(1-D_{2^{'}\bar{1} 3^{'}})=0 .
\end{equation}
Again, it is equivalent to Eq.~\eqref{nMTDIgen} by identifying $a=D_{33^{'}}$ and $b=D_{2^{'}\bar{1} 3^{'}}$.

For the commutator $[a, [a, b]]$, the corresponding TDI solution expressed in terms of the coefficients $q_1, q_{1^{'}}$ reads
\begin{equation}
	\begin{aligned}
		q_{1} &= \alpha =- 2{D_{33'2'\bar 13'}} + {D_{2'\bar 13'33'}} + {D_{33'}} + {D_{2'\bar 13'}} - 1,\\
		q_{1^{'}} &= \beta ={D_{33'33'}} - 2{D_{33'}} + 1,
	\end{aligned}
\end{equation}
which gives rise to
\begin{align}
P_{\Delta_1}=& \left( { - 2{D_{33'2'\bar 13'}} + {D_{2'\bar 13'33'}} + {D_{33'}} + {D_{2'\bar 13'}} - 1} \right){\eta _1} + \left( {{D_{33'33'}} - 2{D_{33'}} + 1} \right){\eta _{1'}}\notag\\
 +& \left( { - {D_{33'33'2'\bar 1}} + 2{D_{33'2'\bar 1}} - {D_{2'\bar 1}}} \right){\eta _2} + \left( { - 2{D_{33'2'\bar 13'3}} + {D_{2'\bar 13'33'3}} + {D_{33'3}} + {D_{2'\bar 13'3}} - {D_3} + {D_{33'33'2'\bar 1}} - 2{D_{33'2'\bar 1}} + {D_{2'\bar 1}}} \right){\eta _{2'}} .
\end{align}

On the other hand, the solution related to $[a, b][a, b]$ is found to be
\begin{equation}
	\begin{aligned}
		q_{1} &= \alpha = {D_{33'2'\bar 13'2'\bar 13'}} + {D_{2'\bar 13'33'}} - {D_{33'2'\bar 13'}} - {D_{2'\bar 13'33'2'\bar 13'}},\\
		q_{1^{'}} &= \beta = {D_{2'\bar 13'33'33'}} + {D_{33'2'\bar 13'}} - {D_{2'\bar 13'33'}} - {D_{33'2'\bar 13'33'}},
	\end{aligned}
\end{equation}
which leads to
\begin{align}
P_{\Delta_2}=& \left( {{D_{33'2'\bar 13'2'\bar 13'}} + {D_{2'\bar 13'33'}} - {D_{33'2'\bar 13'}} - {D_{2'\bar 13'33'2'\bar 13'}}} \right){\eta _1}\notag\\
- & \left( {{D_{2'\bar 13'33'33'2'\bar 1}} + {D_{33'2'\bar 13'2'\bar 1}} - {D_{2'\bar 13'33'2'\bar 1}} - {D_{33'2'\bar 13'33'2'\bar 1}}} \right){\eta _2}\notag\\
 +& \left( {{D_{2'\bar 13'33'33'}} + {D_{33'2'\bar 13'}} - {D_{2'\bar 13'33'}} - {D_{33'2'\bar 13'33'}}} \right){\eta _{1'}} \notag\\
 +& \left[ \begin{array}{c}
\left( {{D_{33'2'\bar 13'2'\bar 13'3}} + {D_{2'\bar 13'33'3}} - {D_{33'2'\bar 13'3}} - {D_{2'\bar 13'33'2'\bar 13'3}}} \right)\\
 + \left( {{D_{2'\bar 13'33'33'2'\bar 1}} + {D_{33'2'\bar 13'2'\bar 1}} - {D_{2'\bar 13'33'2'\bar 1}} - {D_{33'2'\bar 13'33'2'\bar 1}}} \right)
\end{array} \right]{\eta _{2'}}.
\end{align}
We show in Fig.~\ref{fig3} the sensitivity curve, which is again found to be identical to the standard second-generation Beacon combination $P_1^{16}$~\cite{TDI2-1}.

\subsubsection{Sagnac-type solutions}

For the Sagnac-alpha type, it is observed that the coefficients satisfy the following relations
\begin{equation}\label{sagnac}
	{q}_{2}={q}_{1} D_{3}, {q}_{3}=q_{1} D_{31}, {q}_{3^{\prime}}=q_{1^{\prime}} D_{2^{\prime}}, {q}_{2^{\prime}}=q_{1^{\prime}} D_{2^{\prime} 1^{\prime}}.
\end{equation}
It is not difficult to show that two of the above equations are linearly dependent with Eqs.~\eqref{TDIeq}.
To be specific, the relevant equation is obtained from the first line of Eqs.~\eqref{TDIeq} by substituting the above conditions and eliminating $q_2, q_3, q_{2'}$, and $q_{3'}$.
One finds
\begin{equation}
	q_{1}(1-D_{312})+q_{1^{'}}(1-D_{2^{'}1^{'}3^{'}})=0.
\end{equation}
Again, it is equivalent to Eq.~\eqref{nMTDIgen} by identifying $a=D_{312}$ and $b=D_{2^{'}1^{'}3^{'}}$.

For the commutator $[a, [a, b]]$, the corresponding TDI solution expressed in terms of the coefficients $q_1, q_{1^{'}}$ reads
\begin{equation}
	\begin{aligned}
		q_{1} &= \alpha =- 2{D_{3122'1'3'}} + {D_{2'1'3'312}} + {D_{312}} + {D_{2'1'3'}} - 1,\\
		q_{1^{'}} &= \beta ={D_{312312}} - 2{D_{312}} + 1,
	\end{aligned}
\end{equation}
which gives rise to
\begin{align}
\alpha_{\Delta_1}=& \left( { - 2{D_{3122'1'3'}} + {D_{2'1'3'312}} + {D_{312}} + {D_{2'1'3'}} - 1} \right){\eta _1} + \left( { - 2{D_{3122'1'3'3}} + {D_{2'1'3'3123}} + {D_{3123}} + {D_{2'1'3'3}} - {D_3}} \right){\eta _2}\notag\\
 +& \left( { - 2{D_{3122'1'3'31}} + {D_{2'1'3'31231}} + {D_{31231}} + {D_{2'1'3'31}} - {D_{31}}} \right){\eta _3} + \left( {{D_{312312}} - 2{D_{312}} + 1} \right){\eta _{1'}}\notag\\
 +& \left( {{D_{3123122'1'}} - 2{D_{3122'1'}} + {D_{2'1'}}} \right){\eta _{2'}} + \left( {{D_{3123122'}} - 2{D_{3122'}} + {D_{2'}}} \right){\eta _{3'}} .
\end{align}

Also, for the commutator $[a, b][a, b]$, the corresponding TDI solution reads
\begin{equation}
	\begin{aligned}
		q_{1} &= \alpha ={D_{3122'1'3'2'1'3'}} + {D_{2'1'3'312}} - {D_{3122'1'3'}} - {D_{2'1'3'3122'1'3'}},\\
		q_{1^{'}} &= \beta ={D_{2'1'3'312312}} + {D_{3122'1'3'}} - {D_{2'1'3'312}} - {D_{3122'1'3'312}},
	\end{aligned}
\end{equation}
which leads to
\begin{align}
\alpha_{\Delta_2}= & \left( {{D_{3122'1'3'2'1'3'}} + {D_{2'1'3'312}} - {D_{3122'1'3'}} - {D_{2'1'3'3122'1'3'}}} \right){\eta _1}\notag\\
+ & \left( {{D_{3122'1'3'2'1'3'}} + {D_{2'1'3'312}} - {D_{3122'1'3'}} - {D_{2'1'3'3122'1'3'}}} \right){D_3}{\eta _2}\notag\\
 +& \left( {{D_{3122'1'3'2'1'3'}} + {D_{2'1'3'312}} - {D_{3122'1'3'}} - {D_{2'1'3'3122'1'3'}}} \right){D_{31}}{\eta _3}\notag\\
 +& \left( {{D_{2'1'3'312312}} + {D_{3122'1'3'}} - {D_{2'1'3'312}} - {D_{3122'1'3'312}}} \right){\eta _{1'}}\notag\\
 +& \left( {{D_{2'1'3'312312}} + {D_{3122'1'3'}} - {D_{2'1'3'312}} - {D_{3122'1'3'312}}} \right){D_{2'1'}}{\eta _{2'}}\notag\\
 +& \left( {{D_{2'1'3'312312}} + {D_{3122'1'3'}} - {D_{2'1'3'312}} - {D_{3122'1'3'312}}} \right){D_{2'}}{\eta _{3'}}.
\end{align}
We show in Fig.~\ref{fig3} the sensitivity curve, compared against that of the standard second-generation Sagnac combination $\alpha_1^{16}$~\cite{tdi-otto-2015}.

\subsubsection{Fully symmetric Sagnac-type solutions}\label{lastButOneSub}

For the fully-symmetric Sagnac type, we consider the following constraints
\begin{equation}
	q_{3}=-q_{3^{'}}, q_{2}=-q_{2^{'}} ,
\end{equation}
which gives
\begin{equation}
	q_{1}(1-D_{3\bar{1}^{'}2})+q_{1^{'}}(1-D_{2^{'}\bar{1}3^{'}})=0 .
\end{equation}
Again, it is equivalent to Eq.~\eqref{nMTDIgen} by identifying $a=D_{3\bar{1}^{'}2}$ and $b=D_{2^{'}\bar{1}3^{'}}$.

For the commutator $[a, [a, b]]$, the corresponding TDI solution expressed in terms of the coefficients $q_1, q_{1^{'}}$ reads
\begin{equation}
	\begin{aligned}
		q_{1} &= \alpha =- 2{D_{3\bar 1'22'\bar 13'}} + {D_{2'\bar 13'3\bar 1'2}} + {D_{3\bar 1'2}} + {D_{2'\bar 13'}} - 1,\\
		q_{1^{'}} &= \beta ={D_{3\bar 1'23\bar 1'2}} - 2{D_{3\bar 1'2}} + 1,
	\end{aligned}
\end{equation}
which gives rise to
\begin{align}
\zeta_{\Delta_1}=& \left( { - 2{D_{3\bar 1'22'\bar 13'}} + {D_{2'\bar 13'3\bar 1'2}} + {D_{3\bar 1'2}} + {D_{2'\bar 13'}} - 1} \right){\eta _1} + \left( { - {D_{3\bar 1'23\bar 1'22'\bar 1}} + 2{D_{3\bar 1'22'\bar 1}} - {D_{2'\bar 1}}} \right){\eta _2}\notag\\
 -& \left( {2{D_{3\bar 1'22'\bar 13'3\bar 1'}} - {D_{2'\bar 13'3\bar 1'23\bar 1'}} - {D_{3\bar 1'23\bar 1'}} - {D_{2'\bar 13'3\bar 1'}} + {D_{3\bar 1'}}} \right){\eta _3} + \left( {{D_{3\bar 1'23\bar 1'2}} - 2{D_{3\bar 1'2}} + 1} \right){\eta _{1'}}\notag\\
 -& \left( { - {D_{3\bar 1'23\bar 1'22'\bar 1}} + 2{D_{3\bar 1'22'\bar 1}} - {D_{2'\bar 1}}} \right){\eta _{2'}} + \left( {2{D_{3\bar 1'22'\bar 13'3\bar 1'}} - {D_{2'\bar 13'3\bar 1'23\bar 1'}} - {D_{3\bar 1'23\bar 1'}} - {D_{2'\bar 13'3\bar 1'}} + {D_{3\bar 1'}}} \right){\eta _{3'}} .
\end{align}

On the other hand, the commutator $[a, b][a, b]$ gives rise to the solution
\begin{equation}
	\begin{aligned}
		q_{1} &= \alpha ={D_{3\bar 1'22'\bar 13'2'\bar 13'}} + {D_{2'\bar 13'3\bar 1'2}} - {D_{3\bar 1'22'\bar 13'}} - {D_{2'\bar 13'3\bar 1'22'\bar 13'}},\\
		q_{1^{'}} &= \beta = {D_{2'\bar 13'3\bar 1'23\bar 1'2}} + {D_{3\bar 1'22'\bar 13'}} - {D_{2'\bar 13'3\bar 1'2}} - {D_{3\bar 1'22'\bar 13'3\bar 1'2}},
	\end{aligned}
\end{equation}
which leads to
\begin{align}
\zeta_{\Delta_2}=& \left( {{D_{3\bar 1'22'\bar 13'2'\bar 13'}} + {D_{2'\bar 13'3\bar 1'2}} - {D_{3\bar 1'22'\bar 13'}} - {D_{2'\bar 13'3\bar 1'22'\bar 13'}}} \right){\eta _1} \notag\\
- & \left( {{D_{2'\bar 13'3\bar 1'23\bar 1'2}} + {D_{3\bar 1'22'\bar 13'}} - {D_{2'\bar 13'3\bar 1'2}} - {D_{3\bar 1'22'\bar 13'3\bar 1'2}}} \right){D_{2'\bar 1}}{\eta _2}\notag\\
 - & \left( {{D_{3\bar 1'22'\bar 13'2'\bar 13'}} + {D_{2'\bar 13'3\bar 1'2}} - {D_{3\bar 1'22'\bar 13'}} - {D_{2'\bar 13'3\bar 1'22'\bar 13'}}} \right){D_{3\bar 1'}}{\eta _3} \notag\\
 + & \left( {{D_{2'\bar 13'3\bar 1'23\bar 1'2}} + {D_{3\bar 1'22'\bar 13'}} - {D_{2'\bar 13'3\bar 1'2}} - {D_{3\bar 1'22'\bar 13'3\bar 1'2}}} \right){\eta _{1'}}\notag\\
 + & \left( {{D_{2'\bar 13'3\bar 1'23\bar 1'2}} + {D_{3\bar 1'22'\bar 13'}} - {D_{2'\bar 13'3\bar 1'2}} - {D_{3\bar 1'22'\bar 13'3\bar 1'2}}} \right){D_{2'\bar 1}}{\eta _{2'}}\notag\\
 + & \left( {{D_{3\bar 1'22'\bar 13'2'\bar 13'}} + {D_{2'\bar 13'3\bar 1'2}} - {D_{3\bar 1'22'\bar 13'}} - {D_{2'\bar 13'3\bar 1'22'\bar 13'}}} \right){D_{3\bar 1'}}{\eta _{3'}}.
\end{align}
We show in Fig.~\ref{fig3} the sensitivity curve compared against that of the standard second-generation fully-symmetric Sagnac combination $\xi_1^{16}$~\cite{tdi-otto-2015}.

It is worth commenting on the degeneracy observed in the sensitivity curves shown in Fig.~\ref{fig3}.
As a matter of fact, distinct geometric TDI solutions (in the sense that the solutions are not related to each other by any symmetric operation such as spacecraft index permutation) have been found to possess identical sensitivity curves (see, for instance, Figs.~17 and~18 of~\cite{Geo-sister}).
Fig.~\ref{fig3} indicates that the solutions encountered using the algebraic approach essentially inherit such a feature and fall into one of the sensitivity categories pertaining to the same type of TDI combinations.
An effort to classify such a degeneracy in terms of the first module of syzygies has recently been carried out~\cite{Geo-panpan-new}.

\subsubsection{Sagnac-inspired solutions}\label{subNewSol}

Last but not least, for the Sagnac-inspired type recently reported in~\cite{cTDI2gen}, we consider the following constraints
\begin{equation}\label{newConz}
	q_{1}=q_{2}D_{1}, q_{2}=q_{3}D_{2} ,
\end{equation}
which gives
\begin{equation}
	q_{3}\left(1-A\right)+q_{3^{\prime}}\left(1-b\right)=0 .
\end{equation}
It is equivalent to Eq.~\eqref{nMTDIgen} by identifying $A= D_{2133^{\prime} 2^{\prime}}-D_{23^{\prime} 2^{\prime}}+D_{22^{\prime}}-D_{212^{\prime}}+D_{21}=\sum_i a_i$ and $b={D}_{1^{\prime} 3^{\prime} 2^{\prime}}$.

For the commutator $[A, [A, b]]$, the corresponding TDI solution expressed in terms of the coefficients $q_3, q_{3^{'}}$ reads
\begin{equation}\label{newSol1}
	\begin{aligned}
		q_{3} &= \alpha = -2\left( {{D_{2133'2'}} - {D_{23'2'}} + {D_{22'}} - {D_{212'}} + {D_{21}}} \right){D_{2133'2'}}\notag\\
 +& {D_{1'3'2'}}\left( {{D_{2133'2'}} - {D_{23'2'}} + {D_{22'}} - {D_{212'}} + {D_{21}}} \right)\notag\\
 +& \left( {{D_{2133'2'}} - {D_{23'2'}} + {D_{22'}} - {D_{212'}} + {D_{21}}} \right) + {D_{1'3'2'}} - 1,\\
		q_{3^{'}} &= \beta =
\left( {{D_{2133'2'}} - {D_{23'2'}} + {D_{22'}} - {D_{212'}} + {D_{21}}} \right)\left( {{D_{2133'2'}} - {D_{23'2'}} + {D_{22'}} - {D_{212'}} + {D_{21}}} \right)\notag\\
 -& 2\left( {{D_{2133'2'}} - {D_{23'2'}} + {D_{22'}} - {D_{212'}} + {D_{21}}} \right) + 1 ,
	\end{aligned}
\end{equation}
which gives rise to
\begin{align}
&S_{\Delta_1}= \left[ \begin{array}{l}
 - 2\left( {{D_{2133'2'}} - {D_{23'2'}} + {D_{22'}} - {D_{212'}} + {D_{21}}} \right){D_{2133'2'}} + {D_{1'3'2'}}\left( {{D_{2133'2'}} - {D_{23'2'}} + {D_{22'}} - {D_{212'}} + {D_{21}}} \right)\\
 + \left( {{D_{2133'2'}} - {D_{23'2'}} + {D_{22'}} - {D_{212'}} + {D_{21}}} \right) + {D_{1'3'2'}} - 1
\end{array} \right]{D_{21}}{\eta _1}\notag\\
 +& \left[ \begin{array}{l}
 - 2\left( {{D_{2133'2'}} - {D_{23'2'}} + {D_{22'}} - {D_{212'}} + {D_{21}}} \right){D_{2133'2'}} + {D_{1'3'2'}}\left( {{D_{2133'2'}} - {D_{23'2'}} + {D_{22'}} - {D_{212'}} + {D_{21}}} \right)\\
 + \left( {{D_{2133'2'}} - {D_{23'2'}} + {D_{22'}} - {D_{212'}} + {D_{21}}} \right) + {D_{1'3'2'}} - 1
\end{array} \right]{D_2}{\eta _2}\notag\\
 +& \left[ \begin{array}{l}
 - 2\left( {{D_{2133'2'}} - {D_{23'2'}} + {D_{22'}} - {D_{212'}} + {D_{21}}} \right){D_{2133'2'}} + {D_{1'3'2'}}\left( {{D_{2133'2'}} - {D_{23'2'}} + {D_{22'}} - {D_{212'}} + {D_{21}}} \right)\\
 + \left( {{D_{2133'2'}} - {D_{23'2'}} + {D_{22'}} - {D_{212'}} + {D_{21}}} \right) + {D_{1'3'2'}} - 1
\end{array} \right]{\eta _3}\notag\\
 + &\left\{ \begin{array}{c}
\left\{ \begin{array}{c}
\left[ {{{\left( {{D_{2133'2'}} - {D_{23'2'}} + {D_{22'}} - {D_{212'}} + {D_{21}}} \right)}^2} - 2\left( {{D_{2133'2'}} - {D_{23'2'}} + {D_{22'}} - {D_{212'}} + {D_{21}}} \right) + 1} \right]{D_{1'}}\\
 + \left[ \begin{array}{l}
 - 2\left( {{D_{2133'2'}} - {D_{23'2'}} + {D_{22'}} - {D_{212'}} + {D_{21}}} \right){D_{2133'2'}} + {D_{1'3'2'}}\left( {{D_{2133'2'}} - {D_{23'2'}} + {D_{22'}} - {D_{212'}} + {D_{21}}} \right)\\
 + \left( {{D_{2133'2'}} - {D_{23'2'}} + {D_{22'}} - {D_{212'}} + {D_{21}}} \right) + {D_{1'3'2'}} - 1
\end{array} \right]{D_{213}}\\
 - \left[ \begin{array}{l}
 - 2\left( {{D_{2133'2'}} - {D_{23'2'}} + {D_{22'}} - {D_{212'}} + {D_{21}}} \right){D_{2133'2'}} + {D_{1'3'2'}}\left( {{D_{2133'2'}} - {D_{23'2'}} + {D_{22'}} - {D_{212'}} + {D_{21}}} \right)\\
 + \left( {{D_{2133'2'}} - {D_{23'2'}} + {D_{22'}} - {D_{212'}} + {D_{21}}} \right) + {D_{1'3'2'}} - 1
\end{array} \right]{D_2}
\end{array} \right\}{D_{3'}}\\
 + \left[ \begin{array}{l}
 - 2\left( {{D_{2133'2'}} - {D_{23'2'}} + {D_{22'}} - {D_{212'}} + {D_{21}}} \right){D_{2133'2'}} + {D_{1'3'2'}}\left( {{D_{2133'2'}} - {D_{23'2'}} + {D_{22'}} - {D_{212'}} + {D_{21}}} \right)\\
 + \left( {{D_{2133'2'}} - {D_{23'2'}} + {D_{22'}} - {D_{212'}} + {D_{21}}} \right) + {D_{1'3'2'}} - 1
\end{array} \right]{D_2}\\
 - \left[ \begin{array}{l}
 - 2\left( {{D_{2133'2'}} - {D_{23'2'}} + {D_{22'}} - {D_{212'}} + {D_{21}}} \right){D_{2133'2'}} + {D_{1'3'2'}}\left( {{D_{2133'2'}} - {D_{23'2'}} + {D_{22'}} - {D_{212'}} + {D_{21}}} \right)\\
 + \left( {{D_{2133'2'}} - {D_{23'2'}} + {D_{22'}} - {D_{212'}} + {D_{21}}} \right) + {D_{1'3'2'}} - 1
\end{array} \right]{D_{21}}
\end{array} \right\}{\eta _{1'}}\notag\\
 +& \left[ \begin{array}{c}
\left[ {{{\left( {{D_{2133'2'}} - {D_{23'2'}} + {D_{22'}} - {D_{212'}} + {D_{21}}} \right)}^2} - 2\left( {{D_{2133'2'}} - {D_{23'2'}} + {D_{22'}} - {D_{212'}} + {D_{21}}} \right) + 1} \right]{D_{1'}}\\
 + \left[ \begin{array}{l}
 - 2\left( {{D_{2133'2'}} - {D_{23'2'}} + {D_{22'}} - {D_{212'}} + {D_{21}}} \right){D_{2133'2'}} + {D_{1'3'2'}}\left( {{D_{2133'2'}} - {D_{23'2'}} + {D_{22'}} - {D_{212'}} + {D_{21}}} \right)\\
 + \left( {{D_{2133'2'}} - {D_{23'2'}} + {D_{22'}} - {D_{212'}} + {D_{21}}} \right) + {D_{1'3'2'}} - 1
\end{array} \right]{D_{213}}\\
 - \left[ \begin{array}{l}
 - 2\left( {{D_{2133'2'}} - {D_{23'2'}} + {D_{22'}} - {D_{212'}} + {D_{21}}} \right){D_{2133'2'}} + {D_{1'3'2'}}\left( {{D_{2133'2'}} - {D_{23'2'}} + {D_{22'}} - {D_{212'}} + {D_{21}}} \right)\\
 + \left( {{D_{2133'2'}} - {D_{23'2'}} + {D_{22'}} - {D_{212'}} + {D_{21}}} \right) + {D_{1'3'2'}} - 1
\end{array} \right]{D_2}
\end{array} \right]{\eta _{2'}}\notag\\
 +& \left[ {{{\left( {{D_{2133'2'}} - {D_{23'2'}} + {D_{22'}} - {D_{212'}} + {D_{21}}} \right)}^2} - 2\left( {{D_{2133'2'}} - {D_{23'2'}} + {D_{22'}} - {D_{212'}} + {D_{21}}} \right) + 1} \right]{\eta _{3'}},
\end{align}

Similarly, the commutator $[A, b][A, b]$ gives rise to the solution
\begin{equation}\label{newSol2}
	\begin{aligned}
		q_{3} &= \alpha =\left( {{D_{2133'2'1'3'2'1'3'2'}} - {D_{23'2'1'3'2'1'3'2'}} + {D_{22'1'3'2'1'3'2'}} - {D_{212'1'3'2'1'3'2'}} + {D_{211'3'2'1'3'2'}}} \right)\notag\\
 +& \left( {{D_{1'3'2'2133'2'}} - {D_{1'3'2'23'2'}} + {D_{1'3'2'22'}} - {D_{1'3'2'212'}} + {D_{1'3'2'21}}} \right)\notag\\
 -& \left( {{D_{2133'2'1'3'2'}} - {D_{23'2'1'3'2'}} + {D_{22'1'3'2'}} - {D_{212'1'3'2'}} + {D_{211'3'2'}}} \right)\notag\\
 &- \left( {{D_{1'3'2'2133'2'1'3'2'}} - {D_{1'3'2'23'2'1'3'2'}} + {D_{1'3'2'22'1'3'2'}} - {D_{1'3'2'212'1'3'2'}} + {D_{1'3'2'211'3'2'}}} \right),\\
		q_{3^{'}} &= \beta =
{D_{1'3'2'}}{\left( {{D_{2133'2'}} - {D_{23'2'}} + {D_{22'}} - {D_{212'}} + {D_{21}}} \right)^2}\notag\\
 +& \left( {{D_{2133'2'1'3'2'}} - {D_{23'2'1'3'2'}} + {D_{22'1'3'2'}} - {D_{212'1'3'2'}} + {D_{211'3'2'}}} \right)\notag\\
 -& \left( {{D_{1'3'2'2133'2'}} - {D_{1'3'2'23'2'}} + {D_{1'3'2'22'}} - {D_{1'3'2'212'}} + {D_{1'3'2'21}}} \right)\notag\\
 -&\left( {{D_{2133'2'1'3'2'}} - {D_{23'2'1'3'2'}} + {D_{22'1'3'2'}} - {D_{212'1'3'2'}} + {D_{211'3'2'}}} \right)\left( {{D_{2133'2'}} - {D_{23'2'}} + {D_{22'}} - {D_{212'}} + {D_{21}}} \right),
	\end{aligned}
\end{equation}
which leads to
\begin{align}
S_{\Delta_2}=q_1\eta_1+q_2\eta_2+q_3\eta_3+q_{1'}\eta_{1'}+q_{2'}\eta_{2'}+q_{3'}\eta_{3'}.
\end{align}
where
\begin{align}
{q_2} = \left[ \begin{array}{l}
\left( {{D_{2133'2'1'3'2'1'3'2'}} - {D_{23'2'1'3'2'1'3'2'}} + {D_{22'1'3'2'1'3'2'}} - {D_{212'1'3'2'1'3'2'}} + {D_{211'3'2'1'3'2'}}} \right)\\
 + \left( {{D_{1'3'2'2133'2'}} - {D_{1'3'2'23'2'}} + {D_{1'3'2'22'}} - {D_{1'3'2'212'}} + {D_{1'3'2'21}}} \right)\\
 - \left( {{D_{2133'2'1'3'2'}} - {D_{23'2'1'3'2'}} + {D_{22'1'3'2'}} - {D_{212'1'3'2'}} + {D_{211'3'2'}}} \right)\\
 - \left( {{D_{1'3'2'2133'2'1'3'2'}} - {D_{1'3'2'23'2'1'3'2'}} + {D_{1'3'2'22'1'3'2'}} - {D_{1'3'2'212'1'3'2'}} + {D_{1'3'2'211'3'2'}}} \right)
\end{array} \right]{D_2},\\
\end{align}
\begin{align}
{q_1} = \left[ \begin{array}{l}
\left( {{D_{2133'2'1'3'2'1'3'2'}} - {D_{23'2'1'3'2'1'3'2'}} + {D_{22'1'3'2'1'3'2'}} - {D_{212'1'3'2'1'3'2'}} + {D_{211'3'2'1'3'2'}}} \right)\\
 + \left( {{D_{1'3'2'2133'2'}} - {D_{1'3'2'23'2'}} + {D_{1'3'2'22'}} - {D_{1'3'2'212'}} + {D_{1'3'2'21}}} \right)\\
 - \left( {{D_{2133'2'1'3'2'}} - {D_{23'2'1'3'2'}} + {D_{22'1'3'2'}} - {D_{212'1'3'2'}} + {D_{211'3'2'}}} \right)\\
 - \left( {{D_{1'3'2'2133'2'1'3'2'}} - {D_{1'3'2'23'2'1'3'2'}} + {D_{1'3'2'22'1'3'2'}} - {D_{1'3'2'212'1'3'2'}} + {D_{1'3'2'211'3'2'}}} \right)
\end{array} \right]{D_{21}},
\end{align}
\begin{align}
{q_{2'}} =& \left[ \begin{array}{l}
{D_{1'3'2'}}{\left( {{D_{2133'2'}} - {D_{23'2'}} + {D_{22'}} - {D_{212'}} + {D_{21}}} \right)^2}\\
 + \left( {{D_{2133'2'1'3'2'}} - {D_{23'2'1'3'2'}} + {D_{22'1'3'2'}} - {D_{212'1'3'2'}} + {D_{211'3'2'}}} \right)\\
 - \left( {{D_{1'3'2'2133'2'}} - {D_{1'3'2'23'2'}} + {D_{1'3'2'22'}} - {D_{1'3'2'212'}} + {D_{1'3'2'21}}} \right)\\
 - \left( {{D_{2133'2'1'3'2'}} - {D_{23'2'1'3'2'}} + {D_{22'1'3'2'}} - {D_{212'1'3'2'}} + {D_{211'3'2'}}} \right)\left( {{D_{2133'2'}} - {D_{23'2'}} + {D_{22'}} - {D_{212'}} + {D_{21}}} \right)
\end{array} \right]{D_{1'}}\notag\\
 +& \left[ \begin{array}{l}
\left( {{D_{2133'2'1'3'2'1'3'2'}} - {D_{23'2'1'3'2'1'3'2'}} + {D_{22'1'3'2'1'3'2'}} - {D_{212'1'3'2'1'3'2'}} + {D_{211'3'2'1'3'2'}}} \right)\\
 + \left( {{D_{1'3'2'2133'2'}} - {D_{1'3'2'23'2'}} + {D_{1'3'2'22'}} - {D_{1'3'2'212'}} + {D_{1'3'2'21}}} \right)\\
 - \left( {{D_{2133'2'1'3'2'}} - {D_{23'2'1'3'2'}} + {D_{22'1'3'2'}} - {D_{212'1'3'2'}} + {D_{211'3'2'}}} \right)\\
 - \left( {{D_{1'3'2'2133'2'1'3'2'}} - {D_{1'3'2'23'2'1'3'2'}} + {D_{1'3'2'22'1'3'2'}} - {D_{1'3'2'212'1'3'2'}} + {D_{1'3'2'211'3'2'}}} \right)
\end{array} \right]{D_{213}}\notag\\
 -& \left[ \begin{array}{l}
\left( {{D_{2133'2'1'3'2'1'3'2'}} - {D_{23'2'1'3'2'1'3'2'}} + {D_{22'1'3'2'1'3'2'}} - {D_{212'1'3'2'1'3'2'}} + {D_{211'3'2'1'3'2'}}} \right)\\
 + \left( {{D_{1'3'2'2133'2'}} - {D_{1'3'2'23'2'}} + {D_{1'3'2'22'}} - {D_{1'3'2'212'}} + {D_{1'3'2'21}}} \right)\\
 - \left( {{D_{2133'2'1'3'2'}} - {D_{23'2'1'3'2'}} + {D_{22'1'3'2'}} - {D_{212'1'3'2'}} + {D_{211'3'2'}}} \right)\\
 - \left( {{D_{1'3'2'2133'2'1'3'2'}} - {D_{1'3'2'23'2'1'3'2'}} + {D_{1'3'2'22'1'3'2'}} - {D_{1'3'2'212'1'3'2'}} + {D_{1'3'2'211'3'2'}}} \right)
\end{array} \right]{D_2},
\end{align}

\begin{align}
q_{1'}=& \left\{ \begin{array}{c}
\left[ \begin{array}{l}
{D_{1'3'2'}}{\left( {{D_{2133'2'}} - {D_{23'2'}} + {D_{22'}} - {D_{212'}} + {D_{21}}} \right)^2}\\
 + \left( {{D_{2133'2'1'3'2'}} - {D_{23'2'1'3'2'}} + {D_{22'1'3'2'}} - {D_{212'1'3'2'}} + {D_{211'3'2'}}} \right)\\
 - \left( {{D_{1'3'2'2133'2'}} - {D_{1'3'2'23'2'}} + {D_{1'3'2'22'}} - {D_{1'3'2'212'}} + {D_{1'3'2'21}}} \right)\\
 - \left( {{D_{2133'2'1'3'2'}} - {D_{23'2'1'3'2'}} + {D_{22'1'3'2'}} - {D_{212'1'3'2'}} + {D_{211'3'2'}}} \right)\left( {{D_{2133'2'}} - {D_{23'2'}} + {D_{22'}} - {D_{212'}} + {D_{21}}} \right)
\end{array} \right]{D_{1'}}\\
 + \left[ \begin{array}{l}
\left( {{D_{2133'2'1'3'2'1'3'2'}} - {D_{23'2'1'3'2'1'3'2'}} + {D_{22'1'3'2'1'3'2'}} - {D_{212'1'3'2'1'3'2'}} + {D_{211'3'2'1'3'2'}}} \right)\\
 + \left( {{D_{1'3'2'2133'2'}} - {D_{1'3'2'23'2'}} + {D_{1'3'2'22'}} - {D_{1'3'2'212'}} + {D_{1'3'2'21}}} \right)\\
 - \left( {{D_{2133'2'1'3'2'}} - {D_{23'2'1'3'2'}} + {D_{22'1'3'2'}} - {D_{212'1'3'2'}} + {D_{211'3'2'}}} \right)\\
 - \left( {{D_{1'3'2'2133'2'1'3'2'}} - {D_{1'3'2'23'2'1'3'2'}} + {D_{1'3'2'22'1'3'2'}} - {D_{1'3'2'212'1'3'2'}} + {D_{1'3'2'211'3'2'}}} \right)
\end{array} \right]{D_{213}}\\
 - \left[ \begin{array}{l}
\left( {{D_{2133'2'1'3'2'1'3'2'}} - {D_{23'2'1'3'2'1'3'2'}} + {D_{22'1'3'2'1'3'2'}} - {D_{212'1'3'2'1'3'2'}} + {D_{211'3'2'1'3'2'}}} \right)\\
 + \left( {{D_{1'3'2'2133'2'}} - {D_{1'3'2'23'2'}} + {D_{1'3'2'22'}} - {D_{1'3'2'212'}} + {D_{1'3'2'21}}} \right)\\
 - \left( {{D_{2133'2'1'3'2'}} - {D_{23'2'1'3'2'}} + {D_{22'1'3'2'}} - {D_{212'1'3'2'}} + {D_{211'3'2'}}} \right)\\
 - \left( {{D_{1'3'2'2133'2'1'3'2'}} - {D_{1'3'2'23'2'1'3'2'}} + {D_{1'3'2'22'1'3'2'}} - {D_{1'3'2'212'1'3'2'}} + {D_{1'3'2'211'3'2'}}} \right)
\end{array} \right]{D_2}
\end{array} \right\}{D_{3'}}\notag\\
 +& \left[ \begin{array}{l}
\left( {{D_{2133'2'1'3'2'1'3'2'}} - {D_{23'2'1'3'2'1'3'2'}} + {D_{22'1'3'2'1'3'2'}} - {D_{212'1'3'2'1'3'2'}} + {D_{211'3'2'1'3'2'}}} \right)\\
 + \left( {{D_{1'3'2'2133'2'}} - {D_{1'3'2'23'2'}} + {D_{1'3'2'22'}} - {D_{1'3'2'212'}} + {D_{1'3'2'21}}} \right)\\
 - \left( {{D_{2133'2'1'3'2'}} - {D_{23'2'1'3'2'}} + {D_{22'1'3'2'}} - {D_{212'1'3'2'}} + {D_{211'3'2'}}} \right)\\
 - \left( {{D_{1'3'2'2133'2'1'3'2'}} - {D_{1'3'2'23'2'1'3'2'}} + {D_{1'3'2'22'1'3'2'}} - {D_{1'3'2'212'1'3'2'}} + {D_{1'3'2'211'3'2'}}} \right)
\end{array} \right]{D_2}\notag\\
 -& \left[ \begin{array}{l}
\left( {{D_{2133'2'1'3'2'1'3'2'}} - {D_{23'2'1'3'2'1'3'2'}} + {D_{22'1'3'2'1'3'2'}} - {D_{212'1'3'2'1'3'2'}} + {D_{211'3'2'1'3'2'}}} \right)\\
 + \left( {{D_{1'3'2'2133'2'}} - {D_{1'3'2'23'2'}} + {D_{1'3'2'22'}} - {D_{1'3'2'212'}} + {D_{1'3'2'21}}} \right)\\
 - \left( {{D_{2133'2'1'3'2'}} - {D_{23'2'1'3'2'}} + {D_{22'1'3'2'}} - {D_{212'1'3'2'}} + {D_{211'3'2'}}} \right)\\
 - \left( {{D_{1'3'2'2133'2'1'3'2'}} - {D_{1'3'2'23'2'1'3'2'}} + {D_{1'3'2'22'1'3'2'}} - {D_{1'3'2'212'1'3'2'}} + {D_{1'3'2'211'3'2'}}} \right)
\end{array} \right]{D_{21}}.
\end{align}

Now we argue that the solutions Eqs.~\eqref{newSol1} and~\eqref{newSol2} obtained in this subsection are distinct from the ones derived in the existing literature~\cite{D2010, cTDI2gen, Geo-sister}.
The reason is twofold.
First, a second-order commutator of the form Eq.~\eqref{Mi-mi1} cannot be expressed as a commutator given by Eq.~\eqref{sys1} satisfying Eq.~\eqref{permuXY} or their linear summations.
This is because such a commutator must contain an even number of time-translation operators, namely, the difference between a monomial and its arbitrary permutation.
The resulting expression typically possesses the form of Eq.~\eqref{M16V2}, $\Delta=ba^2b-ab^2a$.
In other words, this solution does not belong to any class explored in Refs.~\cite{D2010, cTDI2gen}.
Second, by employing proof by contradiction, one can also show that the solution Eq.~\eqref{newSol1} does not belong to those obtained by the geometric approach~\cite{Geo-sister}.
This is because the geodesic TDI algorithm only enumerates particular solutions that can be associated with a particular light trajectory in the space-time diagram (such as those shown in Fig.~\ref{fig2}).
Such trajectory dictates that for a signal emanated from a given spacecraft, there are only three feasible links (the fourth choice cancels entirely with the preceding link and therefore is not counted)~\cite{Geo-sister}. 
Specifically, according to Fig~\ref{fig1}, the laser noise originated from spacecraft 1, $q_1$, can be propagated along $L_2$ (counter-clockwise $1\to 2$) or $L_{3'}$ (clockwise $1\to 3$), but it will not propagate along $L_1$ (counter-clockwise $3\to 1$) or $L_3$ (counter-clockwise $1\to 2$).
In other words, in a geometric TDI solution, one might encounter factors such as $D_2 q_1$ and $D_{3'} q_1$ but not $D_1q_1$ or $D_3q_1$. 
However, if one assumes the solution given by Eq.~\eqref{newSol1} can be obtained by the geometric TDI approach, it leads to a contradiction.
Since any one of the possible propagations, such as the form $D_2 q_1$, inevitably implies a factor owing to the constraint condition assumed in Eq.~\eqref{newConz}, specifically $D_2 q_2$ in this case, which is excluded by the geometric TDI based on the above arguments.
We also note that the above arguments can be readily borrowed to understand why fully symmetric Sagnac solutions cannot be derived using the geometric TDI approach.

\begin{figure}[h]
\includegraphics[width=0.45\textwidth]{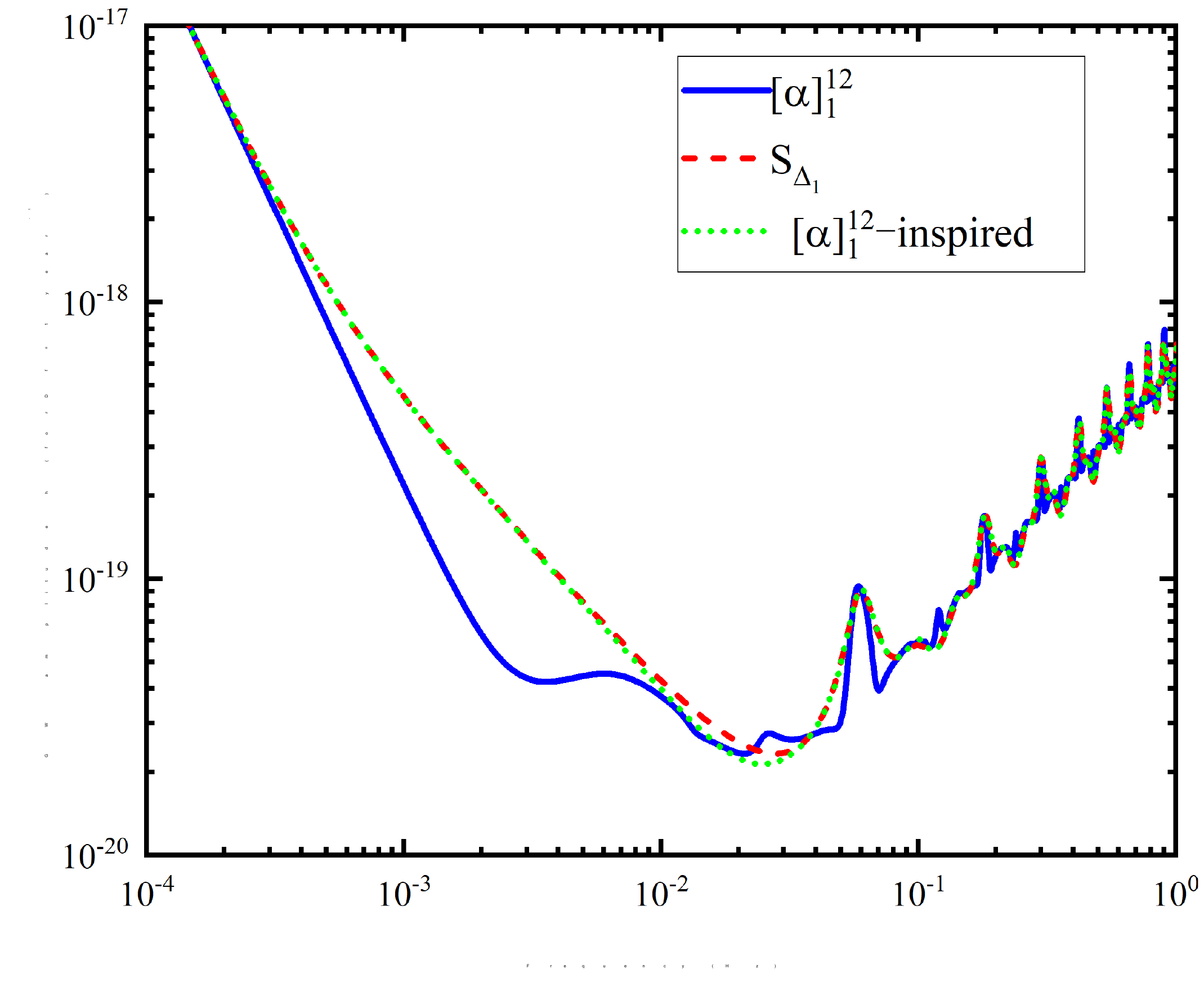}
\caption{\label{fig4} (Color online) The sensitivity curves for Sagnac-inspired solution ($S_{\Delta_1}$) Eq.~\eqref{newSol1} derived in the text, compared against the Sagnac one ($[\alpha]_1^{12}$) and the one ($[\alpha]_1^{12}$-inspired) presented in Eqs.~(74-75) and Fig.~6 of Ref.~\cite{cTDI2gen}.
A solid blue curve presents the solution derived in this study, while the Sagnac one is shown in a dashed red curve, and the one from Ref.~\cite{cTDI2gen} is shown in a dotted green curve.}
\end{figure}

One may illustrate the above conclusion by explicitly evaluating the corresponding sensitivity curve of the solution ($S_{\Delta_1}$) Eq.~\eqref{newSol1} in Fig.~\ref{fig4}, compared against the Sagnac one ($[\alpha]_1^{12}$) and the solution ($[\alpha]_1^{12}$-inspired) presented in Eqs.~(74-75) and Fig.~6 of Ref.~\cite{cTDI2gen}.
The difference between these solutions is clearly observed.

\section{Concluding remarks}\label{sec5}

The present paper involves an attempt to enrich the TDI solutions from an algebraic perspective.
Given the status quo of the ongoing space-borne GW detection programs, laser noise suppression is a relevant subject.
Among others, black hole spectroscopy, which aims at the ringdown-based tests of the no-hair theorem or constraints on modified theories of gravity, calls for unprecedented detector sensitivity.
Even though the second-generation TDI solutions, in principle, provide the means to suppress the laser noise below the noise floor, a better understanding of the TDI solutions and particularly the underlying mathematical foundation, is a pertinent topic.
Unlike the method of exhaustion, which has been most successful in deriving most second-generation TDI solutions, an algebraic approach aims to construct TDI solutions based on a given algorithm.
Besides the mathematical insight, such an approach might become indispensable when the exhaustion approach becomes infeasible due to the computational power.

Following the train of thought first proposed by Dhurandar {\it et al.} in their seminal paper~\cite{D2010}, it turns out that the underlying algebraic {\it solution space} can be significantly expanded further in two distinct directions.
The first possibility~\cite{cTDI2gen} is to replace the original constraint condition (associated with the ``Michelson'' type with one arm dysfunctional) with different ones.
The second choice, proposed in the present manuscript, aims to raise the order of the commutators that furnish the second-generation TDI solutions.
The present study is motivated by the latter, which further explores the relationship between the commutators of the time-displacement operators and second-generation TDI combinations.
We proposed an algebraic algorithm based on second-order commutators, which can be readily generalized to higher-order ones.
To our knowledge, higher-order commutators have not been explored in the literature.
Moreover, the proposed algorithm does not rely on any particular permutation of the time-translation operators nor the vanishing condition of the embedded first-order commutator, which are two indispensable features employed by the remaining algorithms on the market~\cite{D2010, cTDI2gen, Tinto:2022zmf}.
As a result, the proposed algorithm significantly expands the solution space, and specifically, they furnish new solutions that are not covered by any of its predecessors.
In this regard, it paves the way to a better understanding of the relevant module of the non-commutative ring.
Nonetheless, the present algorithm's exhaustive nature is still unclear.
Among others, the solution space is delimited by the specific constraint conditions, and subsequently, only specific subspaces have been explored.
Lastly, as also pointed out by other authors, the major challenge of the algebraic approach for the second-generation TDI is owing to the non-commutative nature of the pertinent polynomial ring.
A general algebraic approach, in this regard, should aim at furnishing the generators of the solution space. 
We plan to address these aspects in further studies.

\section*{Acknowledgements}

We gratefully acknowledge the financial support from Brazilian agencies 
Funda\c{c}\~ao de Amparo \`a Pesquisa do Estado de S\~ao Paulo (FAPESP), 
Funda\c{c}\~ao de Amparo \`a Pesquisa do Estado do Rio de Janeiro (FAPERJ), 
Conselho Nacional de Desenvolvimento Cient\'{\i}fico e Tecnol\'ogico (CNPq), 
and Coordena\c{c}\~ao de Aperfei\c{c}oamento de Pessoal de N\'ivel Superior (CAPES).
This work is also supported by the National Natural Science Foundation of China (Grant No.11925503), the Postdoctoral Science Foundation of China (Grant No.2022M711259), 
Guangdong Major project of Basic and Applied Basic Research (Grant No.2019B030302001), 
Natural Science Foundation of Hubei Province (2021CFB019), 
and the Fundamental Research Funds for the Central Universities, HUST: 2172019kfyRCPY029.

\appendix

\section{The properties of the proposed commutators}\label{appA}

In this Appendix, we elaborate on the proofs of propositions~\ref{prop1} and~\ref{prop2} given in the main text.

Since polynomials are composed of monomials, and due to the relations
\begin{equation}
\begin{aligned}
	[a, b + c] &= [a, b] + [a, c] ,\nonumber\\
	[a+b, c] &= [a, c] + [b, c] ,
\end{aligned}
\end{equation}
for arbitrary monomials $a, b$, and $c$, one only needs to show that propositions~\ref{prop1} and~\ref{prop2} are correct for arbitrary monomials.

A crucial feature of the present scheme is that one also needs to deal with the inverse of the time-delay operator, namely, the time-advance operators that satisfy
\begin{equation}\label{timeAdvD}
	D_{\bar{i}} D_{i} =  D_{i} D_{\bar{i}} = \mathscr{I} ,
\end{equation}
where $\mathscr{I}$ is the identity operator.

When applied to an arbitrary time-dependent variable $\phi(t)$, we have
\begin{equation}
D_{\bar{i}} \phi(t) = \phi (t+L_{i}(t+L_{i})) ,
\end{equation}
where one ignores the second-order terms, which can be expanded to give
\begin{equation}\label{1Dinver}
	\begin{aligned}
		{D}_{\overline{i}} \phi(t) & \simeq \phi\left({t}+{L}_{{i}}\right)+\dot{\phi}\left(t+L_{i}\right)\left(t+L_{i}\right) \dot{L}_{i} \\
		&=\phi\left({t}+{L}_{{i}}\right)+\dot{\phi}\left(t+L_{i}\right) t \dot{L}_{i}+\dot{\phi}\left(t+L_{i}\right) L_{i} \dot{L}_{i}.
	\end{aligned}
\end{equation}

In~\cite{cTDI2gen}, one shows the following generalized form of Eq.~\eqref{sys1}
\begin{equation}\label{sys2}
	\begin{gathered}
		{\left[D_{x_{1} x_{2} \ldots x_{n}}, D_{y_{1} y_{2} \ldots y_{n}}\right] \phi(t)} 
		=\left(\sum_{i=1}^{n} \delta_{{x_i}}L_{x_i} \sum_{j=1}^{n} \delta_{{y_j}}\dot{L}_{y_j}-\sum_{j=1}^{n} \delta_{{y_j}}L_{y_j} \sum_{i=1}^{n} \delta_{{x_i}}\dot{L}_{x_i}\right) 
		\dot{\phi}\left(t-\sum_{k=1}^{n} \delta_{{x_k}}L_{{x_k}}-\sum_{k'=1}^{n} \delta_{y_{k'}}L_{y_{k'}}\right) ,
	\end{gathered}
\end{equation}
where
\begin{equation}
\begin{aligned}
\delta_{x_k} &= \left\{  \begin{matrix}-1&&\mathrm{if}\ k=r\ \mathrm{for\ any}\ r=\mu_1,\cdots,\mu_l\\+1&&\mathrm{otherwise}\end{matrix} \right. ,\\
\delta_{y_{k'}} &= \left\{  \begin{matrix}-1&&\mathrm{if}\ k'=s\ \mathrm{for\ any}\ s=\nu_1,\cdots,\nu_m\\+1&&\mathrm{otherwise}\end{matrix} \right. ,
\end{aligned}
\end{equation}
where one assumes that there are $l$ instances of time-advance operators in $D_{x_{1} x_{2} \dots x_{n}}$ and $m$ instances in $D_{y_{1} y_{2} \dots y_{n}}$.
The subscripts of those time-advance operators are denoted by $\mu_i$ ($=1,\cdots, n$) with $i=1, \cdots, l$ and $\nu_j$ with $j=1, \cdots, m$.

It is not difficult to verify that the vanishing condition Eq.~\eqref{permuXY} remains unchanged when the time-advance operators are considered.
However, two arbitrary monomials, $a$ and $b$, do not necessarily have the same length.
For $n \ne m$,
\begin{equation}
\Delta(n, m) \sim \left[D_{x_{1} x_{2} \ldots x_{n}}, D_{y_{1} y_{2} \ldots y_{m}}\right]
\end{equation}
where ``$\sim$'' indicates the lengths of the two elements of the commutator, Eq.~\eqref{sys2} cannot be utilized straightforwardly.
Nonetheless, for $n=m$, Eq.~\eqref{sys2} leads to
\begin{equation}\label{sys3}
[a, b\Delta(n, m)c]\phi(t) \sim ab\left[D_{x_{1} x_{2} \ldots x_{n}}, D_{y_{1} y_{2} \ldots y_{n}}\right]c\phi(t)- b\left[D_{x_{1} x_{2} \ldots x_{n}}, D_{y_{1} y_{2} \ldots y_{n}}\right] ca\phi(t) = 0 ,
\end{equation}
for arbitrary monomials $a, b$, and $c$.

On the other hand, for the lowest-order commutators
\begin{equation}
	\left[a_1, b_1c_1\right] = [a_1, b_1]c_1+b_1[a_1, c_1] \sim \Delta[1,1]c_1 + b_1\Delta[1,1] ,\nonumber
\end{equation}
and
\begin{equation}
\begin{aligned}
	\left[a_1d_1e_1, b_1c_1\right] &= [a_1, b_1c_1]d_1e_1+a_1[d_1e_1, b_1c_1] =   [a_1, b_1]c_1d_1e_1+b_1[a_1, c_1]d_1e_1+a_1[d_1e_1, b_1c_1] \nonumber\\
	&\sim \Delta[1,1]c_1d_1e_1 + b_1\Delta[1,1]d_1e_1 + a_1\Delta[2,2] ,
\end{aligned}
\end{equation}
where $a_1, b_1$, and $c_1$ are monomials of degree one.
In both cases, the r.h.s. of the equation are written into the summation of some particular $\Delta(i, i)$ multipliers.
The latter is formed by multiplying, from either left and right, by some monomials.
\begin{equation}\label{sys4}
	\left[A, B\right] \sim \sum_i a_i\Delta(n_i, n_i)b_i .
\end{equation}
Although somewhat tedious, it is straightforward to show by {\it mathematical induction} that Eq.~\eqref{sys4} is general.
By combining the above two pieces, one finds that $\Delta_1$ defined by proposition~\ref{prop1} indeed vanishes.

Now, for proposition~\ref{prop2}, one employs Eq.~\eqref{sys4} for both commutators $[A, B]$ and $[C, D]$.
Then it is straightforward to observe that all the terms in the resultant expression are proportional to the form $\dot{L}_x \dot{L}_y$, which is subsequently ignored in the context of second-generation TDI.

The proofs of proposition~\ref{prop3} and corollary~\ref{coro1} and~\ref{coro2} are apparent.

\section{The procedure to map a commutator to the associated TDI solution}\label{appB}

Because a polynomial comprises monomials, one rewrites the monomial as a summation of multipliers in either $(1-a)$ or $(1-b)$.
To achieve this, one performs successive long divisions of individual monomials by either $(1-a)$ or $(1-b)$ from the right.
It is straightforward to show that the last remainder is always ``1''.

For individual monomial $t_n$, its decomposition can be carried out as follows, reminiscent of the procedure discussed in~\cite{cTDI2gen}.
\begin{enumerate}
\item Initiate $\lambda=0$ and $\gamma=0$
\item It is noted that $t_{n}$ ends in either $a$, $\bar{a}$, $b$, or $\bar{b}$, namely, $t_{n}=t_{n-1}a$, $t_{n}=t_{n-1}\bar{a}$, $t_{n}=t_{n-1}b$, or $t_n=t_{n-1}\bar{b}$. 
\begin{itemize}
\item If $t_{n}=t_{n-1}a$, let $\lambda=\lambda-t_{n-1}$;
\item If $t_{n}=t_{n-1}\bar{a}$, let $\lambda=\lambda+t_{n-1}\bar{a}$;
\item If $t_{n}=t_{n-1}b$, let $\gamma=\gamma-t_{n-1}$;
\item If $t_{n}=t_{n-1}\bar{b}$, let $\gamma=\gamma+t_{n-1}\bar{b}$;
\end{itemize}
\item Repeat the above procedure 2 for $t_{n-1}$, until the degree of the monomial vanishes, namely, $t_{0}$.
We note that $t_{0}=1$.
\item We have $t_n=\lambda (1-a) +\gamma (1-b)+1$.
\end{enumerate}
As an example, for the case $t_{4}=ba\bar{b}a$, we have $t_{3}=ba\bar{b}$, $t_{2}=ba$, $t_{1}=b$, $t_{0}=1$. 
So that $\lambda=-t_{3}-t_1=-ba\bar{b}-b$ and $\gamma=t_2\bar{b}+t_0=ba\bar{b}+1$, and subsequently, $ba\bar{b}a=-ba\bar{b}(1-a)+ba\bar{b}(1-b)-b(1-a)-(1-b)+1$.
Since a constructed commutator by the process always involves an even number of monomials with successively flipped signs, the $t_0$ terms are guaranteed to be entirely canceled out.

\section{Sensitivity funtions for the novel combinations}\label{appC}

This Appendix presents the expressions for the averaged response functions of gravitational waves and noise power spectral density for the combinations elaborated in the main text.

For the Michelson-type combination, the noise power spectral density $N_X(u)$ is expressed as
	\begin{equation}\label{cn1}
		{N_X}(u) = 64\left( {3 + \cos \left[ {2u} \right]} \right)\sin {\left[ u \right]^4} \times \frac{{{L^2}s_a^2}}{{{u^2}{c^4}}} + 64\sin {\left[ u \right]^4} \times \frac{{{u^2}s_x^2}}{{{L^2}}},
	\end{equation}
and the gravitational waves averaged response function $R_X(u)$ reads
	\begin{equation}\label{cr1}
		\begin{aligned}
{R_X}(u) = & - \frac{{8\left( {12 - 15\cos \left[ u \right] + 12\cos \left[ {2u} \right]} \right)\sin {{\left[ u \right]}^4}}}{{3{u^2}}} - \frac{{8\sin {{\left[ u \right]}^4}\left( {15\sin \left[ u \right] - 24\cos \left[ u \right]\sin \left[ u \right]} \right)}}{{3{u^3}}} - \frac{{8\sin {{\left[ u \right]}^4}\left( {21\sin \left[ u \right] - 6\sin \left[ {2u} \right]} \right)}}{{3u}}\notag\\
 -& \frac{8}{3}\sin {\left[ u \right]^4}\left[ \begin{array}{l}
 - 5 - \cos \left[ {2u} \right] + 18\cos \left[ {2u} \right]\left( {{\rm{Ci}}\left[ u \right] - 2{\rm{Ci}}\left[ {2u} \right] + {\rm{Ci}}\left[ {3u} \right] + \log \left[ {\frac{4}{3}} \right]} \right)\\
 - 12\left( {{\rm{Ci}}\left[ u \right] - {\rm{Ci}}\left[ {2u} \right] + \log \left[ 2 \right]} \right) + 18\sin \left[ {2u} \right]\left( {{\rm{Si}}\left[ u \right] - 2{\rm{Si}}\left[ {2u} \right] + {\rm{Si}}\left[ {3u} \right]} \right)
\end{array} \right].
		\end{aligned}
	\end{equation}
where $u=2\pi fc/L$ and $f$ is observed frequency, SinIntergral $\mathrm{Si}\left(z\right)=\int_{0}^{z}\mathrm{sin}t/tdt$ and CosIntegral $\mathrm{Ci}\left(z\right)=-\int_{z}^{\infty}\mathrm{cos}t/tdt$.

For the Monitor-type combination, the noise power spectral density $N_E(u)$ is expressed as
	\begin{equation}\label{cn2}
{N_E}(u) = 64\left( {3 + \cos \left[ u \right]} \right)\sin {\left[ {\frac{u}{2}} \right]^4} \times \frac{{{L^2}s_a^2}}{{{u^2}{c^4}}} + 32\left( {3 + 2\cos \left[ u \right]} \right)\sin {\left[ {\frac{u}{2}} \right]^4} \times \frac{{{u^2}s_x^2}}{{{L^2}}},
	\end{equation}
and the gravitational waves averaged response function $R_E(u)$ reads
	\begin{equation}\label{cr2}
		\begin{aligned}
{R_E}(u) =& \frac{{2\left( { - 33 + 45\cos \left[ u \right] - 33\cos \left[ {2u} \right] - 15\cos \left[ {3u} \right]} \right)\sin {{\left[ {\frac{u}{2}} \right]}^4}}}{{3{u^2}}}\notag\\
 +& \frac{{2\sin {{\left[ {\frac{u}{2}} \right]}^4}\left( { - 81\sin \left[ u \right] + 27\sin \left[ {2u} \right] - 3\sin \left[ {3u} \right]} \right)}}{{3u}} + \frac{{2\sin {{\left[ {\frac{u}{2}} \right]}^4}\left( { - 75\sin \left[ u \right] + 33\sin \left[ {2u} \right] + 15\sin \left[ {3u} \right]} \right)}}{{3{u^3}}}\notag\\
 +& \frac{2}{3}\sin {\left[ {\frac{u}{2}} \right]^4}\left[ \begin{array}{l}
20 + 4\cos \left[ u \right] + 168\left( {1 + \cos \left[ u \right]} \right){\rm{Ci}}\left[ u \right] - 240\left( {1 + \cos \left[ u \right]} \right){\rm{Ci}}\left[ {2u} \right] + \\
24\left( {3\left( {1 + \cos \left[ u \right]} \right){\rm{Ci}}\left[ {3u} \right] + \left( {1 + \cos \left[ u \right]} \right)\log \left[ {\frac{{1024}}{{27}}} \right] + 3\sin \left[ u \right]\left( {{\rm{Si}}\left[ u \right] - 2{\rm{Si}}\left[ {2u} \right] + {\rm{Si}}\left[ {3u} \right]} \right)} \right)
\end{array} \right].
		\end{aligned}
	\end{equation}
 
For the Relay-type combination, the noise power spectral density $N_U(u)$ is expressed as
	\begin{align}\label{cn3}
{N_E}(u)= & 256\cos {\left[ {\frac{u}{2}} \right]^2}\left( {5 + 5\cos \left[ u \right] + 2\cos \left[ {2u} \right]} \right)\sin {\left[ {\frac{u}{2}} \right]^4} \times \frac{{{L^2}s_a^2}}{{{u^2}{c^4}}}\notag\\
 +& 128\cos {\left[ {\frac{u}{2}} \right]^2}\left( {4 + 4\cos \left[ u \right] + \cos \left[ {2u} \right]} \right)\sin {\left[ {\frac{u}{2}} \right]^4} \times \frac{{{u^2}s_x^2}}{{{L^2}}},
	\end{align}
and the gravitational waves averaged response function $R_U(u)$ reads
	\begin{equation}\label{cr3}
		\begin{aligned}
{R_U}(u) =& \frac{{4\cos {{\left[ {\frac{u}{2}} \right]}^2}\left( { - 87 + 36\cos \left[ u \right] - 102\cos \left[ {2u} \right] - 48\cos \left[ {3u} \right] - 15\cos \left[ {4u} \right]} \right)\sin {{\left[ {\frac{u}{2}} \right]}^4}}}{{3{u^2}}}\notag\\
+& \frac{{4\cos {{\left[ {\frac{u}{2}} \right]}^2}\sin {{\left[ {\frac{u}{2}} \right]}^4}\left( { - 324\sin \left[ u \right] + 24\left( {\sin \left[ {2u} \right] + \sin \left[ {3u} \right]} \right) - 3\sin \left[ {4u} \right]} \right)}}{{3u}}\notag\\
 +& \frac{{4\cos {{\left[ {\frac{u}{2}} \right]}^2}\sin {{\left[ {\frac{u}{2}} \right]}^4}\left( { - 132\sin \left[ u \right] + 72\sin \left[ {2u} \right] + 48\sin \left[ {3u} \right] + 15\sin \left[ {4u} \right]} \right)}}{{3{u^3}}}\\\notag\\
 + & \frac{4}{3}\cos {\left[ {\frac{u}{2}} \right]^2}\sin {\left[ {\frac{u}{2}} \right]^4}\left\{ \begin{array}{l}
72 + 56\cos \left[ u \right] + 16\cos \left[ {2u} \right]\\
 + 48\left[ \begin{array}{l}
\left( {7 + 6\cos \left[ u \right] - \cos \left[ {2u} \right]} \right){\rm{Ci}}\left[ u \right] - 2\left( {5 + 3\cos \left[ u \right] - 2\cos \left[ {2u} \right]} \right){\rm{Ci}}\left[ {2u} \right] + \cos \left[ {2u} \right]\log \left[ {\frac{{27}}{{16}}} \right]  \\
+\log \left[ {\frac{{1024}}{{27}}} \right] + \cos \left[ u \right]\log \left[ {64} \right] + 6{\rm{Ci}}\left[ {3u} \right]\sin {\left[ u \right]^2}\\
- 3\left( {\sin \left[ u \right] + \sin \left[ {2u} \right]} \right)\left( {{\rm{Si}}\left[ u \right] - 2{\rm{Si}}\left[ {2u} \right] + {\rm{Si}}\left[ {3u} \right]} \right)
\end{array} \right]
\end{array} \right\}.
		\end{aligned}
	\end{equation}
 
For the Beacon-type combination, the noise power spectral density $N_P(u)$ is expressed as
	\begin{equation}\label{cn4}
{N_P}(u) = 256\cos {\left[ {\frac{u}{2}} \right]^2}\left( {3 + \cos \left[ u \right]} \right)\sin {\left[ {\frac{u}{2}} \right]^4} \times \frac{{{L^2}s_a^2}}{{{u^2}{c^4}}} + 32\left( {3 + 2\cos \left[ u \right]} \right)\sin {\left[ {\frac{u}{2}} \right]^2}\sin {\left[ u \right]^2} \times \frac{{{u^2}s_x^2}}{{{L^2}}},
	\end{equation}
and the gravitational waves averaged response function $R_P(u)$ reads
	\begin{equation}\label{cr4}
		\begin{aligned}
{R_P}(u) =& \frac{{8\cos {{\left[ {\frac{u}{2}} \right]}^2}\left( { - 33 + 45\cos \left[ u \right] - 33\cos \left[ {2u} \right] - 15\cos \left[ {3u} \right]} \right)\sin {{\left[ {\frac{u}{2}} \right]}^4}}}{{3{u^2}}}\notag\\
 +& \frac{{8\cos {{\left[ {\frac{u}{2}} \right]}^2}\sin {{\left[ {\frac{u}{2}} \right]}^4}\left( { - 81\sin \left[ u \right] + 27\sin \left[ {2u} \right] - 3\sin \left[ {3u} \right]} \right)}}{{3u}}\notag\\
 +& \frac{{8\cos {{\left[ {\frac{u}{2}} \right]}^2}\sin {{\left[ {\frac{u}{2}} \right]}^4}\left( { - 75\sin \left[ u \right] + 33\sin \left[ {2u} \right] + 15\sin \left[ {3u} \right]} \right)}}{{3{u^3}}}\notag\\
 +& \frac{8}{3}\cos {\left[ {\frac{u}{2}} \right]^2}\sin {\left[ {\frac{u}{2}} \right]^4}\left[ \begin{array}{l}
20 + 4\cos \left[ u \right] + 168\left( {1 + \cos \left[ u \right]} \right){\rm{Ci}}\left[ u \right] - 240\left( {1 + \cos \left[ u \right]} \right){\rm{Ci}}\left[ {2u} \right] + \\
24\left( {3\left( {1 + \cos \left[ u \right]} \right){\rm{Ci}}\left[ {3u} \right] + \left( {1 + \cos \left[ u \right]} \right)\log \left[ {\frac{{1024}}{{27}}} \right] + 3\sin \left[ u \right]\left( {{\rm{Si}}\left[ u \right] - 2{\rm{Si}}\left[ {2u} \right] + {\rm{Si}}\left[ {3u} \right]} \right)} \right)
\end{array} \right].
		\end{aligned}
	\end{equation}
For the Sagnac-$\alpha$-type combination, the noise power spectral density $N_{\alpha}(u)$ is expressed as
	\begin{equation}\label{cn5}
{N_\alpha }(u) = 128{\left( {1 + 2\cos \left[ u \right]} \right)^4}\left( {5 + 4\cos \left[ u \right] + 2\cos \left[ {2u} \right]} \right)\sin {\left[ {\frac{u}{2}} \right]^6} \times \frac{{{L^2}s_a^2}}{{{u^2}{c^4}}} + 96\sin {\left[ {\frac{{3u}}{2}} \right]^4} \times \frac{{{u^2}s_x^2}}{{{L^2}}},
	\end{equation}
and the gravitational waves averaged response function $R_{\alpha}(u)$ reads
	\begin{equation}\label{cr5}
		\begin{aligned}
{R_\alpha }(u) = & \frac{{2\left( { - 72\sin {{\left[ {\frac{u}{2}} \right]}^2} + 180\cos \left[ u \right]\sin {{\left[ {\frac{u}{2}} \right]}^2} - 72\cos \left[ {2u} \right]\sin {{\left[ {\frac{u}{2}} \right]}^2}} \right)\sin {{\left[ {\frac{{3u}}{2}} \right]}^4}}}{{3{u^2}}}\notag\\
 +& \frac{{2\sin {{\left[ {\frac{{3u}}{2}} \right]}^4}\left( { - 180\sin {{\left[ {\frac{u}{2}} \right]}^2}\sin \left[ u \right] + 72\sin {{\left[ {\frac{u}{2}} \right]}^2}\sin \left[ {2u} \right]} \right)}}{{3{u^3}}}\notag\\
 +& \frac{{2\sin {{\left[ {\frac{{3u}}{2}} \right]}^4}\left( { - 204\sin \left[ u \right] + 147\sin \left[ {2u} \right] - 30\sin \left[ {3u} \right]} \right)}}{{3u}}\notag\\
 +& \frac{2}{3}\sin {\left[ {\frac{{3u}}{2}} \right]^4}\left\{ \begin{array}{l}
24\left( {8 + 4\cos \left[ u \right] + 3\cos \left[ {3u} \right]} \right){\rm{Ci}}\left[ u \right] - 48\left( {7 + 2\cos \left[ u \right] + 3\cos \left[ {3u} \right]} \right){\rm{Ci}}\left[ {2u} \right]\\
 + 72\sin {\left[ {\frac{u}{2}} \right]^2} + 32\cos \left[ u \right]\sin {\left[ {\frac{u}{2}} \right]^2} + 16\cos \left[ {2u} \right]\sin {\left[ {\frac{u}{2}} \right]^2}\\
 + 24\left[ \begin{array}{l}
6{\rm{Ci}}\left[ {3u} \right] + 6\log \left[ {\frac{4}{3}} \right] + 3\cos \left[ {3u} \right]\left( {{\rm{Ci}}\left[ {3u} \right] + \log \left[ {\frac{4}{3}} \right]} \right) + \\
\log \left[ 4 \right] + \cos \left[ u \right]\log \left[ {16} \right] + 3\sin \left[ {3u} \right]\left( {{\rm{Si}}\left[ u \right] - 2{\rm{Si}}\left[ {2u} \right] + {\rm{Si}}\left[ {3u} \right]} \right)
\end{array} \right]
\end{array} \right\}.
		\end{aligned}
	\end{equation}

For the fully-symmetric Sagnac type combination, the noise power spectral density $N_{\zeta}(u)$ is expressed as
	\begin{align}\label{cn6}
{N_\zeta }(u)= & 6{e^{ - 3iu}}{\left( { - 1 + {e^{iu}}} \right)^4}\left( { - 1 + 2{e^{iu}}} \right) \times \frac{{{L^2}s_a^2}}{{{u^2}{c^4}}}\notag\\
 +& \left[ {6\left( {5 - 6\cos \left[ u \right] + \cos \left[ {2u} \right] - 2i\left( { - 1 + \cos \left[ u \right]} \right)\sin \left[ u \right]} \right)} \right] \times \frac{{{u^2}s_x^2}}{{{L^2}}},
	\end{align}
and the gravitational waves averaged response function $R_{zeta}(u)$ reads
	\begin{equation}\label{cr6}
		\begin{aligned}
{R_\zeta }(u) = & \frac{1}{{{u^3}}}{{\rm{e}}^{ - \frac{{iu}}{2}}}\left( {\cos \left[ {\frac{u}{2}} \right] + 3i\sin \left[ {\frac{u}{2}} \right]} \right)\sin {\left[ {\frac{u}{2}} \right]^2} \times \notag\\
&\left[ \begin{array}{l}
12{u^3}\left( {2 + 3\cos \left[ u \right]} \right){\rm{Ci}}\left[ u \right] - 24{u^3}\left( {1 + 3\cos \left[ u \right]} \right){\rm{Ci}}\left[ {2u} \right]\\
 + 36{u^3}\cos \left[ u \right]{\rm{Ci}}\left[ {3u} \right] + 36{u^3}\cos \left[ u \right]\log \left[ {\frac{4}{3}} \right] + 12{u^3}\log \left[ 4 \right] + 4{u^3}\sin {\left[ {\frac{u}{2}} \right]^2} + \\
30u\cos \left[ u \right]\sin {\left[ {\frac{u}{2}} \right]^2} + 3{u^2}\sin \left[ u \right] - 3{u^2}\cos \left[ u \right]\sin \left[ u \right] - 30\sin {\left[ {\frac{u}{2}} \right]^2}\sin \left[ u \right]\\
 + 36{u^3}\sin \left[ u \right]{\rm{Si}}\left[ u \right] - 72{u^3}\sin \left[ u \right]{\rm{Si}}\left[ {2u} \right] + 36{u^3}\sin \left[ u \right]{\rm{Si}}\left[ {3u} \right]
\end{array} \right].
		\end{aligned}
	\end{equation}
For all the above cases, the resultant sensitivity function reads
\begin{equation}\label{c7}
	S(u)=\frac{\sqrt{N(u)}}{\sqrt{\frac{2}{5}}\sqrt{R(u)}}.
\end{equation}
The plots shown in Fig.~\ref{fig3} are evaluated using Eqs.~\eqref{cn1}$-$\eqref{c7}.

\bibliographystyle{h-physrev}
\bibliography{reference_TDI}

\begin{thebibliography}{10}

\bibitem{TDI-1}
M.~Tinto and J.~W. Armstrong,
\newblock Physical Review D {\bf 59}, 102003 (1999).

\bibitem{LISA-1}
P.~Amaro-Seoane {\em et~al.},
\newblock arXiv preprint arXiv:1702.00786  (2017).

\bibitem{TianQin-1}
J.~Luo {\em et~al.},
\newblock Classical and Quantum Gravity {\bf 33}, 035010 (2016),
\newblock arXiv: 1512.02076.

\bibitem{Taiji-1}
W.-R. Hu and Y.-L. Wu,
\newblock National Science Review {\bf 4}, 685 (2017).

\bibitem{Overview-1}
M.~Tinto and S.~V. Dhurandhar,
\newblock Living Reviews in Relativity {\bf 24}, 1 (2021).

\bibitem{Geo-TDI-1}
M.~Vallisneri,
\newblock Physical Review D {\bf 72}, 042003 (2005).

\bibitem{TDI1-1}
J.~W. Armstrong, F.~B. Estabrook, and M.~Tinto,
\newblock The Astrophysical Journal {\bf 527}, 814 (1999).

\bibitem{TDI1.5-1}
N.~J. Cornish and R.~W. Hellings,
\newblock Classical and Quantum Gravity {\bf 20}, 4851 (2003).

\bibitem{TA-1}
S.~V. Dhurandhar, K.~R. Nayak, and J.-Y. Vinet,
\newblock Physical Review D {\bf 65}, 102002 (2002).

\bibitem{TA-2}
K.~R. Nayak and J.-Y. Vinet,
\newblock Physical Review D {\bf 70}, 102003 (2004).

\bibitem{groebner}
D.~Cox, J.~Little, and D.~OShea,
\newblock {\em Ideals, varieties, and algorithms: an introduction to
  computational algebraic geometry and commutative algebra} (Springer Science
  \& Business Media, 2013).

\bibitem{TDI2-1}
M.~Tinto, F.~B. Estabrook, and J.~W. Armstrong,
\newblock Physical Review D {\bf 69}, 082001 (2004).

\bibitem{Olaf}
O.~Hartwig and M.~Muratore,
\newblock Phys. Rev. D {\bf 105}, 062006 (2022).

\bibitem{Geo-sister}
P.-P. Wang, W.-L. Qian, Y.-J. Tan, H.-Z. Wu, and C.-G. Shao,
\newblock Phys. Rev. D {\bf 106}, 024003 (2022).

\bibitem{D2010}
S.~V. Dhurandhar, K.~R. Nayak, and J.-Y. Vinet,
\newblock Classical and Quantum Gravity {\bf 27}, 135013 (2010).

\bibitem{cTDI2gen}
Z.-Q. Wu, P.-P. Wang, W.-L. Qian, and C.-G. Shao,
\newblock Phys. Rev. D {\bf 107}, 024042 (2023), arXiv:2210.07801.

\bibitem{TC-1}
M.~Tinto and O.~Hartwig,
\newblock Physical Review D {\bf 98}, 042003 (2018).

\bibitem{D2009}
S.~V. Dhurandhar,
\newblock Journal of Physics: Conference Series {\bf 154}, 012047 (2009).

\bibitem{tdi-otto-2015}
M.~Otto,
\newblock {\em Time-Delay Interferometry Simulations for the Laser
  Interferometer Space Antenna},
\newblock PhD thesis, Leibniz ${\rm{Universit\ddot{a}t}}$ Hannover, 2015.

\bibitem{Geo-panpan-new}
P.-P. e.~a. Wang,
\newblock In preparation  (2023).

\bibitem{Tinto:2022zmf}
M.~Tinto, S.~Dhurandhar, and D.~Malakar,
\newblock Phys. Rev. D {\bf 107}, 082001 (2023), arXiv:2212.05967.

\end{thebibliography}

\end{document}